\documentclass[12pt,a4paper]{article}
\usepackage{amsmath, caption, longtable, paralist, amsfonts, amssymb}
\usepackage[top=1.2in, bottom=1.2in, left=1in, right=1in]{geometry}
\DeclareCaptionStyle{italic}[justification=centering]{labelfont={bf}, textfont={it}, labelsep=colon}
\usepackage{psfrag, epsf}
\usepackage{xcolor}
\usepackage{enumerate}
\usepackage{natbib}
\def\arraystretch{0.75}
\usepackage{url} 
\usepackage{caption}
\usepackage{subcaption}
\usepackage{bbm, dsfont, mathpazo, bm}
\usepackage{comment}
\usepackage{float}
\usepackage{booktabs, rotating} 
\usepackage[final,nopatch=footnote]{microtype}
\usepackage[pdftex,colorlinks=true]{hyperref}
\definecolor{darkblue}{rgb}{0,0,.6}
\hypersetup{citecolor=darkblue,linkcolor=darkblue,urlcolor=darkblue,pdfencoding=auto}

\newcommand{\blind}{0}
\newcommand{\Rlogo}{\protect\includegraphics[height=1.8ex,keepaspectratio]{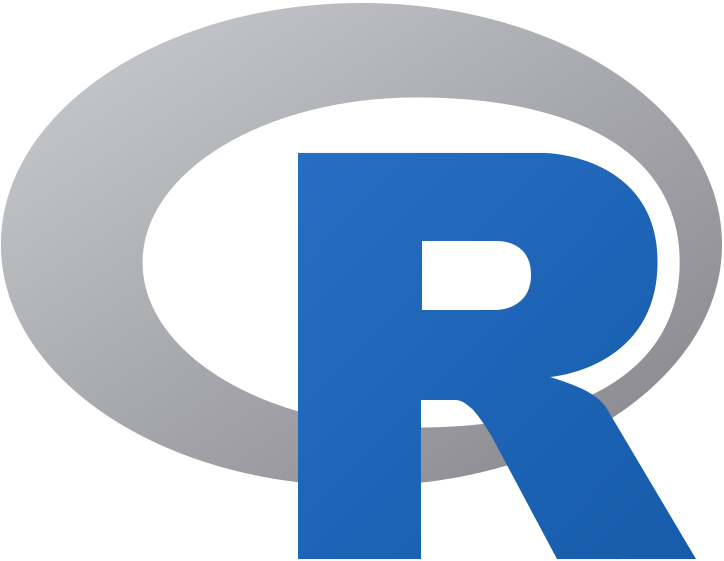}}

\usepackage{mathtools}
\usepackage{graphicx}
\usepackage{multirow}
\newcommand{\norm}[1]{\left\lVert #1 \right\rVert}

\usepackage{algpascal}
\usepackage{textcomp}
\usepackage{amsthm}
\usepackage[export]{adjustbox}
\theoremstyle{remark}
\newtheorem{rem}{Remark}

\usepackage{tikz}

\usepackage{smartdiagram}
\usepackage[utf8]{inputenc}
\usetikzlibrary{arrows,positioning}
\usepackage{makecell} 
\setcellgapes{5pt}

\DeclareMathOperator*{\argmin}{arg\,min}

\newcommand{\X}{\mathcal{X}}
\newcommand{\Y}{\mathcal{Y}}

\graphicspath{{plots/}}
\newsavebox\CBox
\def\textBF#1{\sbox\CBox{#1}\resizebox{\wd\CBox}{\ht\CBox}{\textbf{#1}}}

\usepackage{orcidlink}
\date{}
\AtBeginDocument{}
\usepackage{cleveref}
\usepackage{algorithm}
\usepackage{algpseudocode}

\usepackage{setspace}
\begin{document}
	
\def\spacingset#1{\renewcommand{\baselinestretch}{#1}\small\normalsize} \spacingset{1}

\spacingset{1.4}

\if0\blind
{
\title{\bf Dependence-based fuzzy clustering of functional time series}
\author{\normalsize \'{A}ngel L\'{o}pez-Oriona\orcidlink{0000-0003-1456-7342}\thanks{Postal address: CEMSE Division, Statistics Program, King Abdullah University of Science and Technology (KAUST), Thuwal 23955-6900, Saudi Arabia. E-mail: angel.lopezoriona@kaust.edu.sa}  \ and \ Ying Sun\orcidlink{0000-0001-6703-4270}  \hspace{.2cm}\\  
\normalsize Statistics Program\\ 
\normalsize King Abdullah University of Science and Technology (KAUST)\\ 
\normalsize Thuwal, Saudi Arabia \\ 
\\
\normalsize Han Lin Shang\orcidlink{0000-0003-1769-6430} \\
\normalsize Department of Actuarial Studies and Business Analytics \\
\normalsize Macquarie University \\
\normalsize Sydney, Australia
}
\maketitle
} \fi
	
\if1\blind
{
\title{\bf Dependence-based fuzzy clustering of functional time series}
} \fi
	
\bigskip

\begin{abstract}
Time series clustering is essential in scientific applications, yet methods for functional time series, collections of infinite-dimensional curves treated as random elements in a Hilbert space, remain underdeveloped. This work presents clustering approaches for functional time series that combine the fuzzy $C$-medoids and fuzzy $C$-means procedures with a novel dissimilarity measure tailored for functional data. This dissimilarity is based on an extension of the quantile autocorrelation to the functional context. Our methods effectively groups time series with similar dependence structures, achieving high accuracy and computational efficiency in simulations. The practical utility of the approach is demonstrated through case studies on high-frequency financial stock data and multi-country age-specific mortality improvements.
		
\vspace{.1in}
		
\end{abstract}
	
\newpage

\section{Introduction}\label{sectionintroduction}

Time series clustering partitions unlabeled time series into homogeneous groups based on a chosen similarity measure, aiming to reveal underlying structures without intensive model fitting. Its growing appeal stems from broad applications and challenges like selecting dissimilarity measures or handling dynamic shifts. For comprehensive reviews, see \cite{aghabozorgi2015time} and \cite{maharaj2019time}.

The choice of clustering technique depends on the time series characteristics and goals. Most methods focus on real-valued time series, often identifying distinct shapes using elastic dissimilarity measures like dynamic time warping (DTW) with $K$-means clustering \citep{ikotun2023k}. A key challenge is defining suitable prototypes, though \cite{VA23} shows that $K$-means can be applied without estimating them. Using DTW with $K$-means leads to generalized $K$-means, where standard properties may not hold \citep{vera2021behaviour, VA23}. An alternative is $K$-medoids, which selects a cluster prototype \citep{aghabozorgi2015time}. \cite{vera2021behaviour} used multidimensional scaling to apply standard $K$-means with any dissimilarity matrix. Recent work of \cite{VA23} and \cite{VR24} has advanced robust DTW-based $K$-means methods for time series clustering.

Model-based clustering assumes predefined models for different groups \citep{maharaj1999comparison}. Feature-based methods extract statistical features from each series and cluster them using a dissimilarity measure \citep{lopez2022quantile1, lopez2022quantile2}. For multivariate data, dimensionality reduction, like PCA, is often applied before clustering \citep{li2019multivariate}.

Clustering methods yield either hard (crisp) or soft (fuzzy) outcomes. Hard clustering assigns each series to one group, creating disjoint clusters, while soft clustering allows partial membership across multiple groups, offering more flexibility and richer insights \citep{miyamoto2008algorithms}. Some time series switch between clusters over time \citep{d2009autocorrelation}, which fuzzy clustering effectively captures. It also offers computational efficiency \citep{mcbratney1985application}, works with distribution-free methods, and identifies a “second-best” cluster \citep{everitt2001m}.

In independent and identically distributed (i.i.d.) functional data analysis, clustering is often combined with dimension reduction to address the curse of dimensionality. \cite{TK03} and \cite{GG05} approximated data using fewer basis functions before clustering the coefficients, while \cite{BCJ15} used a functional mixture model for bike-sharing data. \cite{tzeng2018dissimilarity} introduced a dissimilarity measure based on smoothing splines. Further work includes \cite{CPH+23} on mixtures-of-experts for functional data and \cite{BJ11}, later expanded by \cite{JP14b}, on high-dimensional clustering assuming Gaussian principal component scores. For a review, see \cite{chamroukhi2019model} and \cite{ZP23}.

Clustering time series beyond \(\mathbb{R}^n\) is less explored. Some methods exist for integer-valued \citep{cerqueti2022ingarch} and categorical \citep{garcia2015framework} sequences, often in biological analysis. Functional time series, indexed infinite-dimensional curves in Hilbert space \citep{hormann2012functional}, are used for intraday stock returns \citep{kokoszka2012functional} and mortality rates \citep{hyndman2009forecasting}. While i.i.d. functional data clustering is well-studied, functional time series clustering remains underdeveloped.

Clustering functional time series has been studied only by \cite{van2021similarity} and \cite{tang2022clustering} (the former in a nonstationary context), highlighting the need for specialized algorithms. Challenges include: 
\begin{inparaenum}
\item[(i)] classical methods do not handle functional data \citep{ramsay2005}, 
\item[(ii)] difficulties defining serial dependence \citep{valencia2019kendall}, and 
\item[(iii)] high-dimensionality \citep{kim2024projection}, which may explain the limited research.
\end{inparaenum}

The high dimensionality and complexity of temporal data make the fuzzy approach effective for clustering functional time series. Switching behavior, common in real-valued time series \citep{d2009autocorrelation}, also occurs in functional time series but is more complex due to intricate serial dependence structures in high-dimensional curves. This paper introduces soft clustering procedures for functional time series to:
\begin{inparaenum}
\item[(i)] group series with similar underlying processes,
\item[(ii)] achieve high clustering accuracy across diverse datasets, and
\item[(iii)] efficiently compute the clustering partition. We combine traditional fuzzy clustering with a novel dissimilarity measure tailored for functional data.
\end{inparaenum}

We define a dissimilarity measure for functional time series. Since our objective is to cluster the series according to their generating processes, the metric is based on certain quantities, giving information about the serial dependence structures. The extracted features rely on a measure of serial dependence, which can be considered as a functional extension of the standard quantile autocorrelation. Thus, the considered measure inherits the nice properties of the latter quantity in a clustering context, as the ability to identify complex dependence patterns that standard correlation-based methods are unable to detect or the low computational complexity \citep{lafuente2016clustering}.

The proposed distance is used with fuzzy $C$-medoids \citep{krishnapuram1999fuzzy, lopez2022spatial} and fuzzy $C$-means \citep{d2009autocorrelation} algorithms to assign gradual memberships to clusters. The main difference is that $C$-medoids uses time series as prototypes (medoids), while $C$-means uses virtual prototypes (centroids). We evaluate the clustering techniques through simulation experiments with various stochastic processes and compare them to other methods. The techniques are applied to intraday stock returns and age-specific mortality improvement rates. 

The paper is organized as follows: Section~\ref{sectiondistance} introduces a serial dependence measure for functional processes, used to construct a dissimilarity for functional time series. Section~\ref{sectionfuzzyalgorithm} designs clustering algorithms combining the proposed dissimilarity with existing methods. Section~\ref{sectionsimulations} evaluates the performance through numerical experiments, comparing competing procedures. Section~\ref{sectionapplication} applies the methods to intraday stock returns and age-specific mortality improvement rates. Finally, Section~\ref{sectionconclusions} presents conclusions and potential extensions.
	
\section{Measuring dissimilarity between functional time series}\label{sectiondistance}
	
Let $\{\mathcal{X}_t, t \in \mathbb{Z}\}$ be a strictly stationary stochastic process formed by real functional random variables taking values in the Hilbert space $H=L^2(\mathcal{I})$, with $\mathcal{I}=[0,1]$, equipped with the inner product $\langle f, g\rangle_H=\int_0^1 f(u) g(u) d u$, for two functions $f, g \in H$, with the corresponding norm $\|\cdot\|_H=\sqrt{\langle\cdot, \cdot\rangle_H}$. A natural measure of serial dependence for functional time series is introduced and used to define a dissimilarity measure.
	
\subsection{A measure of serial dependence for functional time series}\label{subsectionmeasuring}
	
Let us denote $F_{\mathcal{X}(u)}$ as the marginal cumulative distribution function of the strictly stationary process $\{\mathcal{X}_t(u), t \in \mathbb{Z}$, $u \in \mathcal{I}\subset \mathbb{R}\}$, and $q_\tau$ as the corresponding functional quantile of level $\tau \in (0,1)$, that is, 
\begin{equation}\label{fquantile}
q_\tau(u)=\inf \{x \in \mathbb{R}: F_{\mathcal{X}(u)}(x) \ge \tau\} \quad \text{ for}\quad u \in \mathcal{I}.
\end{equation}
	
For a given lag, $l \in \mathbb{Z}$, two quantile levels, $\left(\tau, \tau^{\prime}\right) \in(0,1)^2$, and two thresholds, $(\beta, \beta') \in [0, 1]^2$, consider the covariance between the indicator functions $\mathds{1}_\beta(\mathcal{X}_t, q_\tau)$ and $\mathds{1}_{\beta'}(\mathcal{X}_{t+l}, q_{\tau'})$ given by
\begin{equation*}
\gamma(\tau, \tau', l, \beta, \beta')=\operatorname{Cov}\Big[\mathds{1}_\beta(\mathcal{X}_t, q_\tau),\mathds{1}_{\beta'}(\mathcal{X}_{t+l}, q_{\tau'})\Big], 
\end{equation*}
where 	
\begin{equation*}
\mathds{1}_\kappa(f, g)=\begin{cases} 
		1   &\text{if} \quad  \mathcal{L}(A) \le \kappa,       \\ 
		0   &\text{if} \quad  \mathcal{L}(A) > \kappa,
\end{cases} 
\end{equation*}
for two functions $f, g \in H$ and $\kappa \in [0,1]$, being 	
\begin{equation*}
A = \{u \in \mathcal{I}: f(u) \le g(u)\},
\end{equation*}
and $\mathcal{L}$ the Lebesgue measure. 
	
As the quantity $\gamma(\tau, \tau', l, \beta, \beta')$ can be seen as an extension of the classical quantile autocovariance to the functional case, we term this quantity the \textit{functional quantile autocovariance of lag $l$ and thresholds $\beta$ and $\beta'$ for levels $\tau$ and $\tau'$}. Note that $\gamma(\tau, \tau', l, \beta, \beta')$ can be expressed as 	
\begin{equation*}
\gamma(\tau, \tau', l, \beta, \beta')=P\big(\mathcal{L}(A_\tau^t)\le \beta,\mathcal{L}(A^{t+l}_{\tau'})\le \beta' \big)-P\big(\mathcal{L}(A^{t}_{\tau})\le \beta\big)P\big(\mathcal{L}(A^{t+l}_{\tau'})\le \beta'\big),
\end{equation*}
where $A^t_{\alpha}=\{u \in \mathcal{I}: \mathcal{X}_t(u) \le q_\alpha(u)\}$ for $\alpha \in [0, 1]$, and that this quantity can be normalized to the interval $[-1, 1]$ by considering 	
\begin{equation*}	
\rho(\tau, \tau', l, \beta, \beta')=\frac{\gamma(\tau, \tau', l, \beta, \beta')}{\Big[P\big(\mathcal{L}(A^{t}_{\tau})\le \beta\big)P\big(\mathcal{L}(A^{t+l}_{\tau'})\le \beta'\big)\big[1-P\big(\mathcal{L}(A^{t}_{\tau})\le \beta\big)\big]\big[1-P\big(\mathcal{L}(A^{t+l}_{\tau'})\le \beta'\big)\big]\Big]^{1/2}},
\end{equation*}
which we refer to as the \textit{functional quantile autocorrelation} (FQA) \textit{of lag $l$ and thresholds $\beta$ and $\beta'$ for levels $\tau$ and $\tau'$}.  Note that $\gamma(\tau, \tau', l, \beta, \beta')$ and $\rho(\tau, \tau', l, \beta, \beta')$ take zero value for an i.i.d. functional process.
	
In practice, the previously defined quantities must be estimated from a $T$-length realization of the functional stochastic process $\{\mathcal{X}_t, t \in \mathbb{Z}\}$, $\boldsymbol{\mathcal{X}}=(\X_1, \X_2, \ldots, \X_T)$, often referred to as \textit{functional time series}. 
We will assume that each function in the realization is observed in a common set of $p$ evenly spaced points, $\mathfrak{U}=\{u_1, u_2, \ldots, u_p\} \subset [0,1]$, with $u_1=0$ and $u_p=1$. Therefore, the information in the realization $\boldsymbol{\mathcal{X}}$ can be expressed through the matrix
\begin{equation*}
L =  
\begin{pmatrix}
\X_1(u_1) &   \X_1(u_2)             & \cdots    & \X_1(u_p)\\
\X_2(u_1) &   \X_2(u_2)             & \cdots    & \X_2(u_p)\\
\vdots      &   \vdots                  & \ddots    & \vdots\\
\X_T(u_1) &   \X_T(u_2)             & \cdots    & \X_T(u_p)
\end{pmatrix},
\end{equation*}
where $\X_{t}(u_j)$ is the value of the function $\X_t$ at the point $u_j$, $t=1,\ldots,T$, $j=1,\ldots,p$. Note that estimates of $\gamma(\tau, \tau', l, \beta, \beta')$ and $\rho(\tau, \tau', l, \beta, \beta')$ can be computed by estimating the probabilities $P\big(\mathcal{L}(A_\tau^t) \le \beta\big)$ and $P\big(\mathcal{L}(A_\tau^t)\le \beta,\mathcal{L}(A^{t+l}_{\tau'})\le \beta' \big)$ in a natural way as 	
\begin{align*}
\widehat{P}\big(\mathcal{L}(A_\tau^t) \le \beta\big) &=\frac{1}{T}\sum_{i=1}^{T}\mathbb{I}\bigg(\frac{\# \widehat{A}_{\tau}^{i}}{p} \le \beta\bigg), \\ 
\widehat{P}\big(\mathcal{L}(A_\tau^t)\le \beta,\mathcal{L}(A^{t+l}_{\tau'})\le \beta' \big) &=\frac{1}{T-l}\sum_{i=1}^{T-l}\mathbb{I}\bigg(\frac{\# \widehat{A}_{\tau}^{i}}{p} \le \beta\bigg)\mathbb{I}\bigg(\frac{\# \widehat{A}_{\tau'}^{i+l}}{p} \le \beta'\bigg),
\end{align*}
respectively, where $\mathbb{I}(\cdot)$ denotes the binary indicator function, the notation $\#$ is used to indicate the cardinal of a set, and $\widehat{A}_{\alpha}^{k}=\{u \in \mathfrak{U}: \X_k(u) \le \widehat{q}_\alpha(u)\}$ for $k=1, \ldots, T$ and $\alpha \in (0,1)$, being $\widehat{q}_\alpha(\cdot)$ the standard estimate of the $\alpha$\textsuperscript{th} quantile of a real functional random variable as defined in~\eqref{fquantile}. The corresponding estimates for $\gamma(\tau, \tau', l, \beta, \beta')$ and $\rho(\tau, \tau', l, \beta, \beta')$ are denoted as $\widehat{\gamma}(\tau, \tau', l, \beta, \beta')$ and $\widehat{\rho}(\tau, \tau', l, \beta, \beta')$, respectively.

Some comments on the thresholds $\beta$ and $\beta'$ in the introduced quantities are provided below. 	
\begin{rem}\label{remthresholds}
\textit{On the importance of thresholds and their choice}. The thresholds \(\beta\) and \(\beta'\) in \(\gamma(\tau, \tau', l, \beta, \beta')\) evaluate serial dependence by considering the "proportion of time" functions \(\X_t\) and \(\X_{t+l}\) are above or below quantile curves. This is natural in the functional setting, as serial dependence can vary across different regions of the function’s domain. The quantile levels \(\tau\) and \(\tau'\) determine the thresholds, allowing for the expression \(\gamma(\tau, \tau', l, \tau, \tau')=P\big(\mathcal{L}(A_\tau^t)\le \tau,\mathcal{L}(A^{t+l}_{\tau'})\le \tau' \big)-0.25\) and \(\rho(\tau, \tau', l, \tau, \tau')=\frac{\gamma(\tau, \tau', l, \tau, \tau')}{0.25}\). This extends the classical quantile autocovariance function \citep{lafuente2016clustering} to the functional setting.
\end{rem}
	
\subsection{A distance measure for functional time series}\label{subsectiondissimilarity}
	
Let $\mathcal{X}_t^{(1)}$ and $\mathcal{X}_t^{(2)}$ be two strictly stationary functional processes. A simple mechanism to evaluate dissimilarity between them relies on comparing their corresponding FQA-based features on common sets of lags, quantile levels, and thresholds. In particular, for a given set of $L$ lags, $\mathcal{L}=\{l_1, l_2, \ldots, l_L\}$, a set of $P$ quantile levels, $\mathcal{T}=\{\tau_1, \tau_2, \ldots, \tau_P\}$, and a collection of $B$ thresholds, $\mathcal{B}=\{\beta_1, \beta_2, \ldots, \beta_B\}$, we define a distance $d_{\text{FQA}}\in [0,1]$ as 
\begin{equation*}
d_{\text{FQA}}\big(\mathcal{X}_t^{(1)}, \mathcal{X}_t^{(2)}\big)=\frac{1}{4LP^2B^2}\sum_{k=1}^{L}\sum_{i_1=1}^{P}\sum_{i_2=1}^{P}\sum_{j_1=1}^{B}\sum_{j_2=1}^{B}\Big[\rho^{(1)}(\tau_{i_1}, \tau_{i_2}, l_k, \beta_{j_1}, \beta_{j_2})-\rho^{(2)}(\tau_{i_1}, \tau_{i_2}, l_k, \beta_{j_1}, \beta_{j_2})\Big]^{2},
\end{equation*}
where we use the superscripts \textsuperscript{(1)} and \textsuperscript{(2)} to indicate that the corresponding features refer to the processes $\mathcal{X}_t^{(1)}$ and $\mathcal{X}_t^{(2)}$, respectively.

In practice, $d_{\text{FQA}}$ must be computed from realizations $\boldsymbol{\mathcal{X}}^{(1)}=(\X_1^{(1)}, \X_2^{(1)}, \ldots, \X_{T_1}^{(1)})$ and $\boldsymbol{\mathcal{X}}^{(2)}=(\X_1^{(2)}, \X_2^{(2)}, \ldots, \X_{T_2}^{(2)})$ of the stochastic processes $\mathcal{X}_t^{(1)}$ and $\mathcal{X}_t^{(2)}$, respectively, by means of 
\begin{equation*}
\widehat{d}_{\text{FQA}}\big(\mathcal{X}_t^{(1)}, \mathcal{X}_t^{(2)}\big)= \frac{1}{4LP^2B^2}\sum_{k=1}^{L}\sum_{i_1=1}^{P}\sum_{i_2=1}^{P}\sum_{j_1=1}^{B}\sum_{j_2=1}^{B}\Big[\widehat{\rho}^{(1)}(\tau_{i_1}, \tau_{i_2}, l_k, \beta_{j_1}, \beta_{j_2})-\widehat{\rho}^{(2)}(\tau_{i_1}, \tau_{i_2}, l_k, \beta_{j_1}, \beta_{j_2})\Big]^{2}.
\end{equation*}
	
A remark concerning the proposed dissimilarity is provided below.

\begin{rem}\label{remfeaturemetric}
\textit{{Feature-based dissimilarities for functional time series}}. The metric \(\widehat{d}_{\text{FQA}}\) is a feature-based dissimilarity \citep{fulcher2018feature}, where identifying latent clustering depends on selecting appropriate features. Once suitable features are chosen, these metrics can compare series of unequal lengths. Calculating features that capture serial dependence in functional time series can be computationally demanding, especially when using measures like autocorrelation \citep{kokoszka2017inference}. In contrast, FQA-derived features are more efficient, relying on simple estimations and pairwise comparisons.
\end{rem}

\section{Soft clustering procedures for functional time series}\label{sectionfuzzyalgorithm}

We describe two fuzzy clustering algorithms for functional time series based on the metric \(\widehat{d}_{\text{FQA}}\): a fuzzy $C$-medoids model (Section \ref{subsectionfuzzyalgorithm1}) and a fuzzy $C$-means model (Section 2 of the Supplement). We also discuss strategies for selecting the necessary hyperparameters.

\subsection{Using \texorpdfstring{$\widehat{d}_{\text{FQA}}$}{} in combination with the fuzzy \texorpdfstring{$C$}{C}-medoids model}\label{subsectionfuzzyalgorithm1}

Let us consider a set of $n$ functional time series, $\mathbb{S}=\{\boldsymbol{\mathcal{X}}^{(1)}, \boldsymbol{\mathcal{X}}^{(2)}, \ldots, \boldsymbol{\mathcal{X}}^{(n)}\}$, where the $i$\textsuperscript{th} series has length $T_i$. We perform fuzzy clustering on the elements of $\mathbb{S}$ in such a way that the series generated from similar stochastic processes are grouped together. To this aim, we propose considering the fuzzy $C$-medoids clustering model introduced by \cite{krishnapuram1999fuzzy} in combination with the metric $\widehat{d}_{\text{FQA}}$ proposed in Section~\ref{subsectiondissimilarity}. Thus, the goal is to determine the subset of $\mathbb{S}$ of size $C$, $\widetilde{\mathbb{S}}=\{\widetilde{\boldsymbol{\mathcal{X}}}^{(1)}, \widetilde{\boldsymbol{\mathcal{X}}}^{(2)}, \ldots, \widetilde{\boldsymbol{\mathcal{X}}}^{(C)}\}$, the elements of which are usually referred to as medoids, and the $n \times C$ matrix of fuzzy coefficients, $\boldsymbol U=(u_{ic})$, with $i=1, \ldots, n$ and $c=1, \ldots, C$, leading to the solution of the minimization problem 
\begin{equation}\label{fcm}
\min_{\widetilde{\mathbb{S}}, \boldsymbol U}\sum_{i=1}^{n}\sum_{c=1}^{C}u_{ic}^m\times \widehat{d}_{\text{FQA}}(i, c),  \quad \text{ with respect to} \quad \sum_{c=1}^{C}u_{ic}=1, \ u_{ic} \ge 0,
\end{equation}
where $\widehat{d}_{\text{FQA}}(i,c)=\widehat{d}_{\text{FQA}}\big(\boldsymbol{\mathcal{X}}^{(i)}, \widetilde{\boldsymbol{\mathcal{X}}}^{(c)}\big)$, $u_{ic} \in [0,1]$ represents the membership degree of the $i$\textsuperscript{th} series in the $c$\textsuperscript{th} cluster, and $m > 1$ is the so-called fuzziness parameter, which regulates the fuzziness of the partition. For $m=1$, we obtain the hard version of the model, so the solution takes the form $u_{ic}=1$ if the $i$\textsuperscript{th} series pertains to cluster $c$ and $u_{ic}=0$ otherwise. As the value of $m$ increases, the boundaries between groups get softer, which results in a fuzzier clustering partition. {Some variants of the above fuzzy $C$-medoids model are given by fuzzy clustering with multi-medoids \citep{mei2011fuzzy}, and convex fuzzy $C$-medoids \citep{pinheiro2020convex}.}

The Lagrangian multipliers method can be used to solve the constrained optimization problem in~\eqref{fcm}, leading to an iterative procedure that alternately optimizes the membership degrees and the medoids. The iterative solutions for the membership degrees \citep{hoppner1999fuzzy}  are given by 
\begin{equation}\label{updatemem}
u_{ic}=\Bigg\{\sum_{c'=1}^{C}\Bigg[\frac{\widehat{d}_{\text{FQA}}(i, c)}{\widehat{d}_{\text{FQA}}(i, c')}\Bigg]^{\frac{1}{m-1}}\Bigg\}^{-1}. 
\end{equation}
	
Once the membership degrees are obtained through~\eqref{updatemem}, the $C$ series minimizing the objective function in \eqref{fcm} are selected as the new medoids. For each $c \in \{1,\ldots,C\}$, the index $j_c$ satisfying	
\begin{equation*}
j_c = \argmin_{1 \leq j \leq n} \sum_{i=1}^{n} u_{ic}^m \times \widehat{d}_{\text{FQA}}\big(\boldsymbol{\mathcal{X}}^{(i)}, \boldsymbol{\mathcal{X}}^{(j)}\big),
\end{equation*} 
is obtained. The above two-step procedure is repeated until there is no change in the medoids or a maximum number of iterations is reached. A description of the corresponding clustering algorithm is given in Algorithm~1 in the Supplement. 

In the numerical analyses, the maximum number of iterations was set to 100,000. One key benefit of the proposed model is its ability to identify prototypes (medoids) that represent each group’s structural information. These medoids can replace the original series in data analysis, reducing computational complexity. Additionally, fuzzy $C$-medoids approaches are generally more robust to noise and outliers than prototype-based methods like fuzzy $C$-means \citep{pinheiro2020convex}.

\subsection{{Selection of hyperparameters}}\label{subsectionhs}

{Selection of the five hyperparameters ($C$, $m$, $\mathcal{L}$, $\mathcal{T}$ and $\mathcal{B}$) involved in the fuzzy clustering procedures based on $\widehat{d}_{\text{FQA}}$ is addressed.}

\subsubsection{{Selection of the set \texorpdfstring{$\mathcal{L}$}{L}}}\label{subsubsectionhs1}

The set of lags ($\mathcal{L}$) can be determined using the selection procedure proposed by \cite{lopez2023hard} (see Section 3.4), adapted to the functional setting. This approach identifies significant lags for each functional time series in the collection and establishes a maximum lag across all series. The testing relies on a $t$-test based on the distance correlation proposed by \cite{szekely2013distance}, implemented through the \textit{dcor.xy()} function in the \Rlogo\ package \textbf{fda.usc} \citep{fda.usc}. Details on this selection procedure are provided in Section~3.1 of the Supplement.

\subsubsection{{Selection of the sets \texorpdfstring{$\mathcal{T}$}{T} and \texorpdfstring{$\mathcal{B}$}{B}}}\label{subsubsectionhs2}

{For the collection of quantile levels ($\mathcal{T}$), we suggest using three quantiles of levels 0.1, 0.5, and 0.9, and, with respect to the set of thresholds ($\mathcal{B}$), we suggest considering $\tau_{i_1}=\beta_{j_1}$ and $\tau_{i_2}=\beta_{j_2}$ (refer to Remark~\ref{remthresholds}) in the framework of Section~\ref{subsectiondissimilarity}. The corresponding explanations are given in Section~3.2 of the Supplement.} 

\subsubsection{{Selection of \texorpdfstring{$C$}{C} and \texorpdfstring{$m$}{m}}}\label{subsubsectionhs3}

{Given fixed values of $\mathcal{L}$, $\mathcal{T}$, and $\mathcal{B}$, the number of clusters ($C$) and the fuzziness parameter ($m$) can be determined by executing the clustering procedure for various combinations of $(C, m)$ and selecting the pair that minimizes the validity index proposed by \citeauthor{xie1991validity} \citeyearpar{xie1991validity}. A detailed explanation of this selection procedure is provided in Section 3.3 of the Supplement.} 

\section{Monte-Carlo simulation studies}\label{sectionsimulations}

Several simulations were conducted to evaluate the proposed clustering techniques in various settings. Competing methods are described, the simulation mechanism and performance measures are outlined, and the results are discussed.


\subsection{Competing methods}
		
The proposed clustering techniques were compared with analogous clustering approaches based on competing metrics. {First, we considered two metrics constructed using the (estimated) functional autocorrelation function proposed by \cite{kokoszka2017inference} and the (estimated) functional spherical autocorrelation function proposed by \cite{yeh2023functional}. For the time series $\boldsymbol{\mathcal{X}}=(\X_1, \X_2, \ldots, \X_T)$ and a lag $l \in \mathbb{Z}$, the first autocorrelation is defined as}

{
\begin{equation*}
\widehat{\rho}_{\text{FACF}}(l)=\frac{\sqrt{\int_0^1\int_0^1 \widehat{C}_l^2(u, v) d u d v}}{\int_0^1 \widehat{C}_0(u, u) d u},
\end{equation*}
where, for $h \in \mathbb{Z}$,
}

{\begin{equation*}
\widehat{C}_h(u, v)=\frac{1}{T} \sum_{i=1}^{T-h}\big[\X_i(u)-\overline{\X}_T(u)\big]\big[\X_{i+h}(v)-\overline{\X}_T(v)\big],
\end{equation*}
with $\overline{\X}_T(w)=\frac{1}{T} \sum_{i=1}^T \X_i(w)$, $w \in [0,1]$, is the estimate of the autocovariance kernel for lag~$h$ defined as $C_h(u, v)=\operatorname{Cov}\big[\X_t(u), \X_{t+h}(v)\big]$ for the underlying functional stochastic process $\{\mathcal{X}_t, t \in \mathbb{Z}\}$. The functional spherical autocorrelation at lag $l$ is defined as}

{\begin{equation*}
\widehat{\rho}_{\text{FSACF}}(l)=\frac{1}{T} \sum_{i=1}^{T-l}\left\langle S\left(\X_i-\widehat{\mu}_{\mathcal{X}_t}\right), S\left(\X_{i+l}-\widehat{\mu}_{\mathcal{X}_t}\right)\right\rangle_H,
\end{equation*}
{where, for $f, g\in H$, $S(f)=\frac{f}{\|f\|_H}$, the inner product $\langle f, g\rangle_H$ and the norm $\|f\|_H$ were defined in Section~\ref{sectiondistance}, and $\widehat{\mu}_{\mathcal{X}_t}$ is the estimate of the so-called spatial median of the process $\{\mathcal{X}_t, t \in \mathbb{Z}\}$ according to the estimation procedure described in \cite{gervini2008robust}. The spatial median of a functional random variable $\X \in H$ is defined as $\mu=\argmin_{\Y \in H} \mathbb{E}\big[\|\X-\Y\|_H-\|\X\|_H\big]$.}}

{Based on previous considerations, given the functional time series $\boldsymbol{\mathcal{X}}^{(1)}$ and $\boldsymbol{\mathcal{X}}^{(2)}$ and the collection of lags $\mathcal{L}=\{l_1, \ldots, l_L\}$, we define two (estimated) distances $\widehat{d}_{\text{FACF}}$ and $\widehat{d}_{\text{FSACF}}$ as
\begin{align*}
\widehat{d}_{\text{FACF}}\big(\mathcal{X}_t^{(1)}, \mathcal{X}_t^{(2)}\big)&= \frac{1}{4L}\sum_{k=1}^{L}\Big[\widehat{\rho}^{(1)}_{\text{FACF}}(l_k)-\widehat{\rho}^{(2)}_{\text{FACF}}(l_k)\Big]^2, \\
\widehat{d}_{\text{FSACF}}\big(\mathcal{X}_t^{(1)}, \mathcal{X}_t^{(2)}\big)&= \frac{1}{4L}\sum_{k=1}^{L}\Big[\widehat{\rho}^{(1)}_{\text{FSACF}}(l_k)-\widehat{\rho}^{(2)}_{\text{FSACF}}(l_k)\Big]^{2},
\end{align*}
where the superscripts \textsuperscript{(1)} and \textsuperscript{(2)} indicate that the corresponding estimates are obtained from the realizations $\boldsymbol{\mathcal{X}}^{(1)}$ and $\boldsymbol{\mathcal{X}}^{(2)}$, respectively. In this paper, practical implementation of the dissimilarities $\widehat{d}_{\text{FACF}}$ and $\widehat{d}_{\text{FSACF}}$ was carried out using the function \textit{obtain\_FACF()} of the \Rlogo\ package \textbf{fdaACF} \citep{fdaacf}, and the functions \textit{fSACF()} and \textit{fSACF\_test()} of the \Rlogo\ package \textbf{FTSgof} \citep{ftsgof}, respectively.}

In addition to the previous two dissimilarities, we considered two metrics relying on the (estimated) functional version of the Kendall correlation coefficient proposed by \cite{valencia2019kendall} adapted to the temporal setting. For the functional time series, $\boldsymbol{\mathcal{X}}$ and a lag $l \in \mathbb{Z}$, this quantity is defined as
\begin{equation*}
\widehat{\rho}_{\text{K}}(l)=\binom{T-l}{2}^{-1} \sum_{i=1}^{T-l}\sum_{j>i}\big[2 \mathbb{I}(\X_i \prec \X_{j})\mathbb{I}(\X_{i+l} \prec \X_{j+l})+2 \mathbb{I}(\X_j \prec \X_{i})\mathbb{I}(\X_{j+l} \prec \X_{i+l})\big]-1,
\end{equation*}
where $\prec$ denotes one of the following pre-orders for two continuous functions $f$ and $g$ defined in~$\mathcal{I}$: 
\begin{equation}\label{preorders}
\begin{split}
&	f \prec_{m} g \equiv \max_{u \in \mathcal{I}} f(u) < \max _{u \in \mathcal{I}} g(u), \\
&	f \prec_{i} g  \equiv \int_0^1\big(g(u)-f(u)\big) du > 0.
\end{split}
\end{equation}
		
Given the realizations $\boldsymbol{\mathcal{X}}^{(1)}$ and $\boldsymbol{\mathcal{X}}^{(2)}$ and the collection of lags $\mathcal{L}$, we define a (estimated) distance $\widehat{d}_{\text{K}}$ as
\begin{equation*}
\widehat{d}_{\text{K}}\big(\mathcal{X}_t^{(1)}, \mathcal{X}_t^{(2)}\big)= \frac{1}{4L}\sum_{k=1}^{L}\Big[\widehat{\rho}^{(1)}_{\text{K}}(l_k)-\widehat{\rho}^{(2)}_{\text{K}}(l_k)\Big]^2.
\end{equation*}
where the superscripts have the same meaning as in the definition of $\widehat{d}_{\text{FACF}}$ and $\widehat{d}_{\text{FSACF}}$. The pre-orders $\prec_m$ and $\prec_i$ defined in~\eqref{preorders} give rise to two particular forms of the metric $\widehat{d}_{\text{K}}$, which are denoted as $\widehat{d}_{\text{K}_\text{m}}$ and $\widehat{d}_{\text{K}_\text{i}}$, respectively. 

The proposed approach was also compared with the model-based clustering procedure introduced by \cite{tang2022clustering}, which relies on a multilevel functional data model applied to functional panel data. Although this technique cannot be directly associated with a distance measure, we will denote it as $\widehat{d}_{\text{TSY}}$, which constitutes a hard clustering approach that automatically determines the number of clusters using a specific measure of within-cluster dispersion. In any case, $\widehat{d}_{\text{TSY}}$ can still be compared with the proposed approach in a reasonable manner in some of the simulation experiments (refer to Section~\ref{subsubsection1as}). The approach $\widehat{d}_{\text{TSY}}$ is implemented via the function \textit{mftsc()} of the \Rlogo\ package \textbf{ftsa} \citep{ftsa}, which was employed for the analyses carried out.

The dissimilarities $\widehat{d}_{\text{FACF}}$ and $\widehat{d}_{\text{FSACF}}$ are very natural competitors for the distance $\widehat{d}_{\text{FQA}}$, as they are both based on well-established autocorrelation measures for functional time series. The metrics $\widehat{d}_{\text{K}_\text{m}}$ and $\widehat{d}_{\text{K}_\text{i}}$ rely on a direct extension to the temporal setting of a Kendall correlation coefficient between functional data. This dependence measure employs functional preorders to quantify the association between two sets of curves in a robust manner, and its effectiveness has been demonstrated in several real applications \citep{valencia2019kendall}. Thus, $\widehat{d}_{\text{K}_\text{m}}$ and $\widehat{d}_{\text{K}_\text{i}}$ are robust dissimilarities that can detect differences between functional time series in terms of dependence structures, as $\widehat{d}_{\text{FQA}}$. Therefore, it is also reasonable to compare these dissimilarities with the proposed distance. Comparisons between the clustering procedures based on the five metrics mentioned above provide insight into the ability of the corresponding dependence measures to discriminate between underlying dynamic patterns. In contrast, the model-based procedure $\widehat{d}_{\text{TSY}}$ follows a completely different approach (e.g., it is not based on fuzzy $C$-medoids or fuzzy $C$-means), but its inclusion as a benchmark method in the comparisons is justified by the fact that $\widehat{d}_{\text{TSY}}$ is, to the best of our knowledge, the only method available in the literature for clustering multiple functional time series.

\subsection{Simulation settings and results}\label{subsectionedr}

The performance of the clustering approaches based on $\widehat{d}_{\text{FQA}}$ was evaluated through an extensive simulation study with different functional processes. Two evaluation systems were used: the first assesses the methods' ability to detect well-separated groups in four-cluster scenarios, while the second evaluates their ability to identify two clusters and detect the fuzzy behavior of an ambiguous time series. In particular, a specific threshold must be fixed to determine when a given time series is assumed to belong to a given cluster.

\subsubsection{Scenarios with well-defined clusters}\label{subsubsection1as}
	
We considered two scenarios containing four clusters characterized by specific data-generating processes, denoted by $\mathcal{C}_1$, $\mathcal{C}_2$, $\mathcal{C}_3$, and $\mathcal{C}_4$, which define the true partition. Each one of the groups includes five $T$-length functional time series, thus resulting in a set of 20 $T$-length functional time series. The corresponding data-generating processes are provided below for both scenarios.
\begin{description}
\item[Scenario 1 -- Linear model] Fuzzy clustering of functional time series generated from linear functional autoregressive (FAR) models \citep{bosq2000linear}. Let $\{\mathcal{X}_t, t \in \mathbb{Z}\}$ be a functional stochastic process following the FAR(2) model given by
\begin{equation*}
\mathcal{X}_t(u)=\int_{0}^{1} \Gamma_1(u, v) \mathcal{X}_{t-1}(v) d v+\int_{0}^{1} \Gamma_2(u, v) \mathcal{X}_{t-2}(v) d v+\epsilon_t(u),
\end{equation*}
where $\Gamma_1(u, v)=c_1 \exp[-c_2(u^2+v^2)]$, $\Gamma_2(u, v)=c_3 \exp[-c_4(u^2+v^2)]$, with $c_1, c_2, c_3, c_4 \in~\mathbb{R}$, and $\epsilon_t(u)$ is an independent Brownian motion over $[0, 1]$ with zero mean and variance $1/T$. The previous scenario is motivated by the first simulation setting in \citet[][Section~3]{wang2023nonlinear}. The vector of coefficients $(c_1, c_2, c_3, c_4)$ is set as $(-0.3, 0.1, 0, 0)$, $(0.3, 0.3, 0, 0)$, $(-0.4, 0.5, -0.3, 0.5)$, and $(0.4, 0.7, 0.3, 0.7)$ for clusters $\mathcal{C}_1$, $\mathcal{C}_2$, $\mathcal{C}_3$, and $\mathcal{C}_4$, respectively.
\item[Scenario~2 -- Nonlinear model] Fuzzy clustering of functional time series generated from nonlinear models. We considered a complex setting with two different types of processes: 
\begin{inparaenum}
\item[(i)] nonlinear FAR processes, and 
\item[(ii)] functional GARCH (fGARCH) models proposed by \cite{aue2017functional}. 
\end{inparaenum}
The specific form of the considered processes is described below for each one of the classes.
\begin{itemize}
\item Let $\{\mathcal{X}_t, t \in \mathbb{Z}\}$ be a functional stochastic process following the nonlinear FAR(1) model given by
\begin{equation*}
\mathcal{X}_t(u)=0.75 \times \int_0^1 \Gamma_1(u, v) \mathcal{X}_{t-1}(v) d v \times \exp \bigg(\int_0^1 \Gamma_1(u, v) \mathcal{X}_{t-1}(v) d v\bigg)+\epsilon_t(u),
\end{equation*}
where $\Gamma_1(u, v)$ and $\epsilon_t(u)$ are defined as previously indicated. This scenario is motivated by the third simulation setting in \citet[][Section 3]{wang2023nonlinear}.
\item Let $\{\mathcal{X}_t, t \in \mathbb{Z}\}$ be a functional stochastic process following the fGARCH$(1, 1)$ model given by the equations
\begin{align*}
\mathcal{X}_t(u) & =\sigma_t(u) \epsilon_t(u), \\
\sigma_t^2(u) & =\delta(u)+\alpha \mathcal{X}_{t-1}^2(u)+\beta \sigma_{t-1}^2(u),
\end{align*}
where $\{\epsilon_t, t \in \mathbb{Z}\}$ is a process formed by i.i.d. random functions and $\delta$ is a nonnegative function. The integral operators $\alpha$ and $\beta$ are defined as $(\alpha f)(u)=\int_0^1\alpha(u, v) f(v) d v$, $(\beta f)(u)=\int_0^1\beta(u, v) f(v) d v$, with $u, v \in [0,1]$, $f \in H=L^2(\mathcal{I})$, being $\alpha(u, v)$ and $\beta(u, v)$ elements of the Hilbert space $H^2=L^2(\mathcal{I}^2)$. We consider $\delta(u)=0.01$, $\alpha(u, v)=cu(1-u) v(1-v)$, $\beta(u, v)=\alpha(u, v)$, where $c$ is a constant, and     
\begin{equation*}
\epsilon_t(u)=\frac{\sqrt{\ln 2}}{2^{200 u}} B_t\left(\frac{2^{400 u}}{\ln 2}\right),
\end{equation*}
with the process $\{B_t, t \in \mathbb{Z}\}$ being formed by i.i.d. standard Brownian motions. The previous setting is motivated by the simulations experiments performed in \citet[][Section 4]{aue2017functional}. 
\end{itemize}
\end{description} 
In Scenario~2, clusters $\mathcal{C}_1$ and $\mathcal{C}_2$ are associated with nonlinear FAR(1) models with vectors of coefficients $(c_1, c_2)=(0.5, 0.5)$ and $(c_1, c_2)=(0.9, 0.5)$, respectively, while clusters $\mathcal{C}_3$ and $\mathcal{C}_4$ are associated with fGARCH(1, 1) models with $c=14$ and $c=15$, respectively. As an exploratory step, we carried out a metric two-dimensional scaling (2DS) based on the distance $\widehat{d}_{\text{FQA}}$. This analysis is provided in Section 4 of the Supplement. 

The numerical experiments were performed in the following way. Given a scenario, five functional time series were simulated from each generating process, thus giving rise to a dataset of 20 series to be subject to the clustering techniques. Two values of the series length were used, namely $T \in \{200, 600\}$. A number of $p=100$ discretization points evenly spaced on $[0,1]$ was employed in all cases. Five different values were considered for the fuzziness parameter, namely $m \in \{1.2, 1.4, 1.6, 1.8, 2\}$. Similar grids have been employed in related works \citep{d2009autocorrelation, lopez2021quantile, lopez2023two}. Note that, when $m=1$, we obtain the crisp version of the corresponding clustering models, while very large values of $m$ result in a substantial degree of overlap between groups. 

Two hundred simulations were performed for a given scenario and specific values for $m$ and $T$. In each trial, the fuzzy $C$-medoids and fuzzy $C$-means methods based on $\widehat{d}_{\text{FQA}}$, $\widehat{d}_{\text{FACF}}$, $\widehat{d}_{\text{FSACF}}$, $\widehat{d}_{\text{K}_\text{m}}$, and $\widehat{d}_{\text{K}_\text{i}}$ were executed using the specific value of $m$ as input. The parameter $C$ was set as $C=4$ (the number of different generating processes). The computation of the three metrics was carried out by using the collections of lags $\mathcal{L}=\{1, 2\}$ and $\mathcal{L}=\{1\}$ in Scenarios 1 and 2, respectively, hence considering the maximum number of defining lags in each case. In addition, for the computation of $\widehat{d}_{\text{FQA}}$, the collection $\mathcal{T}$ was set as $\{0.1, 0.5, 0.9\}$ (see Section~\ref{subsectionhs}). Moreover, we also performed the analyses described in this section by considering different choices for the sets mentioned above, but no significant improvements were found.

For each metric and clustering algorithm, at each simulation trial, the following two-step mechanism was considered in order to circumvent the well-known problem of reaching local minima: 
\begin{enumerate}
\item[1)] Executing the corresponding procedure by using 200 initializations for the collection of medoids (fuzzy $C$-medoids) or the membership matrix (fuzzy $C$-means) and recording the final matrix of membership degrees and the final prototypes for each initialization. 
\item[2)] Choosing the clustering solution (among the 200 computed ones) associated with the minimum value of the objective function, refer to~\eqref{fcm}.  
\end{enumerate}

Some studies on fuzzy time series clustering use multiple random initializations and select the one that optimizes the objective function or a given quality index \citep{lopez2022spatial, lopez2023two}, thus avoiding the need for a specific initialization procedure for the medoids or membership matrix.

Clustering accuracy was assessed using the fuzzy extensions of the Adjusted Rand Index (ARIF) and the Jaccard Index (JIF) \citep{campello2007fuzzy}, which compare the true hard partition with the fuzzy solution. ARIF and JIF range from $[-1, 1]$ and $[0, 1]$, respectively, with values closer to 1 indicating better accuracy.

The method $\widehat{d}_{\text{TSY}}$ cannot be directly compared with those remaining. On the one hand, since $\widehat{d}_{\text{TSY}}$ constitutes a hard clustering approach and the current scenarios consider underlying hard partitions, this method is expected to naturally lead to higher values of ARIF and JIF than the alternative techniques. On the other hand, $\widehat{d}_{\text{TSY}}$ automatically determines the number of clusters in its first stage, which is a disadvantage with respect to the approaches based on $\widehat{d}_{\text{FQA}}$, $\widehat{d}_{\text{FACF}}$, $\widehat{d}_{\text{FSACF}}$, $\widehat{d}_{\text{K}_\text{m}}$, and $\widehat{d}_{\text{K}_\text{i}}$, since the latter methods use the true value $C=4$ as input. To carry out fair comparisons, we considered only the trials in which $\widehat{d}_{\text{TSY}}$ selected the true number of clusters ($C=4$). 

The mean values of the clustering quality indexes for the fuzzy $C$-medoids procedures based on the five dissimilarity measures are presented in the upper and lower parts of Table~\ref{table1as} for Scenarios 1 and 2, respectively. Note that for larger values of $m$, the methods exhibit lower clustering accuracy. This is expected, as increasing the fuzziness parameter spreads the membership degrees more evenly across groups, leading to a fuzzier partition and worse performance measures. To rigorously compare the five distances, pairwise paired $t$-tests were conducted using the 200 simulation trials.

\begin{table}[!htb]
\centering
\tabcolsep 0.123in
\caption{\small {Average ARIF and JIF (in brackets) for different fuzzy $C$-medoids methods in the crisp scenarios. The highest scores are highlighted in bold. An asterisk indicates that the corresponding method is more effective than the ones without an asterisk for a significance level of 0.01.}}\label{table1as}
{\begin{tabular}{@{}llllllll@{}}
\toprule 
 Model & $T$ &  $m$ & $\widehat{d}_{\text{FQA}}$  & $\widehat{d}_{\text{FACF}}$ & $\widehat{d}_{\text{FSACF}}$  & $\widehat{d}_{\text{K}_\text{m}}$ & $\widehat{d}_{\text{K}_\text{i}}$  \\\midrule
Linear & 200 & $1.2$   & 0.90 (0.86) & 0.48 (0.45) & 0.41 (0.39) & 0.90 (0.87)  & \textBF{0.94}$^*$ (\textBF{0.91}$^*$)   \\
 & & $1.4$ & 0.80 (0.74) & 0.45 (0.42) & 0.37 (0.37) & 0.88 (0.83) & \textBF{0.91}$^*$ (\textBF{0.88}$^*$)      \\
&   & $1.6$   & 0.70 (0.63) & 0.41 (0.39) & 0.33 (0.34) & 0.84 (0.79) & \textBF{0.88}$^*$ (\textBF{0.83}$^*$)     \\
&  & $1.8$   & 0.61 (0.55) & 0.37 (0.37) & 0.30 (0.32) & 0.79 (0.73) & \textBF{0.83}$^*$ (\textBF{0.78}$^*$)     \\
&  & $2.0$   & 0.52 (0.48) & 0.34 (0.35) & 0.26 (0.31) & 0.74 (0.68) & \textBF{0.79}$^*$ (\textBF{0.72}$^*$)    \\
\cmidrule{2-8}
& 600 & $1.2$   & \textBF{0.99}$^*$ (\textBF{0.99}$^*$)  & 0.61 (0.55) & 0.54 (0.49) & \textBF{0.99}$^*$ (\textBF{0.99}$^*$) &  \textBF{0.99}$^*$ (\textBF{0.99}$^*$)  \\
&  & $1.4$  & 0.98$^*$ (0.96$^*$) & 0.57 (0.51) & 0.50 (0.46) & \textBF{0.99}$^*$ (\textBF{0.99}$^*$) &  \textBF{0.99}$^*$ (\textBF{0.99}$^*$)   \\
&  & $1.6$   & 0.93 (0.89) & 0.53 (0.48) & 0.46 (0.43) & 0.97$^*$ (0.96$^*$) &  \textBF{0.98}$^*$ (\textBF{0.97}$^*$)      \\
&  & $1.8$ & 0.86 (0.80) & 0.49 (0.46) & 0.42 (0.41) & 0.94$^*$ (0.91$^*$) &  \textBF{0.95}$^*$ (\textBF{0.93}$^*$)    \\
&  & $2.0$  & 0.78 (0.71) & 0.46 (0.43) & 0.39 (0.39) & \textBF{0.91}$^*$ (0.86$^*$) &  \textBF{0.91}$^*$ (\textBF{0.88}$^*$)     \\\midrule
Nonlinear & 200 & $1.2$   & \textBF{0.86}$^*$ (\textBF{0.81}$^*$)  &  0.49 (0.45) &  0.54 (0.48) & 0.56 (0.50) & 0.59 (0.53)    \\
& & $1.4$ & \textBF{0.78}$^*$ (\textBF{0.72}$^*$) & 0.46 (0.43) & 0.50 (0.45) & 0.52 (0.47) & 0.55 (0.50)    \\
& & $1.6$   & \textBF{0.69}$^*$ (\textBF{0.62}$^*$) & 0.43 (0.41) & 0.46 (0.43) &  0.48 (0.44)  & 0.53 (0.57)    \\
& & $1.8$   &  \textBF{0.61}$^*$ (\textBF{0.55}$^*$) & 0.40 (0.39) & 0.43 (0.40) & 0.45 (0.41) & 0.50 (0.44)    \\ & & $2.0$   & \textBF{0.53}$^*$ (\textBF{0.48}$^*$) & 0.38 (0.37) & 0.40 (0.38) & 0.41 (0.39) & 0.46 (0.41)    \\
\cmidrule{2-8}
& 600 & $1.2$  & \textBF{0.99}$^*$ (\textBF{0.98}$^*$) & 0.98$^*$ (0.97$^*$) & 0.67 (0.60) & 0.70 (0.64) & 0.68 (0.62)    \\
& & $1.4$  & \textBF{0.96}$^*$ (\textBF{0.93}$^*$) & 0.94$^*$ (0.92$^*$) & 0.64 (0.57) &  0.66 (0.60) & 0.65 (0.60)     \\
& & $1.6$   & 0.90$^*$ (0.85$^*$) &\textBF{0.91}$^*$ (\textBF{0.87}$^*$) & 0.61 (0.55) & 0.62 (0.56) & 0.63 (0.59)   \\
& & $1.8$ & 0.83 (0.77) & \textBF{0.87}$^*$ (\textBF{0.82}$^*$) & 0.58 (0.52) & 0.57 (0.52) & 0.60 (0.56) \\
& & $2.0$  & 0.76 (0.68) &\textBF{0.83}$^*$ (\textBF{0.77}$^*$) & 0.54 (0.49) & 0.53 (0.49) & 0.57 (0.54)    \\ 				
\bottomrule
\end{tabular}}
\end{table}

In each setting, the alternative hypotheses stated that the mean ARIF (JIF) value of a given method is greater than the mean ARIF (JIF) value of its counterpart. Bonferroni corrections were applied to the set of $p$-values associated with each scenario, clustering quality index, and value of $T$. An asterisk was incorporated in Table~\ref{table1as} if the corresponding method was: 
\begin{inparaenum}
\item[(i)] no significantly worse than any other method for a significance level of 0.01, and 
\item[(ii)] significantly more effective than those without an asterisk for a significance level of 0.01.
\end{inparaenum}
The method $\widehat{d}_{\text{TSY}}$ achieves substantially lower scores than the remaining approaches in most cases. {In Scenario 1, $\widehat{d}_{\text{TSY}}$ reaches an ARIF (JIF) of 0.23 (0.30) and 0.27 (0.35) for $T=200$ and $T=600$, respectively. In Scenario 2, it achieves an ARIF (JIF) of 0.36 (0.35) and 0.41 (0.44) for $T=200$ and $T=600$, respectively. These scores suggest that this technique is not suitable for clustering when the differences are driven by the dependence structures of the functional time series. For this reason, the results associated with $\widehat{d}_{\text{TSY}}$ are omitted in subsequent analyses.}

According to Table~\ref{table1as}, the metrics based on Kendall autocorrelations, $\widehat{d}_{\text{K}_\text{m}}$ and $\widehat{d}_{\text{K}_\text{i}}$, achieve a higher clustering accuracy than the proposed dissimilarity in Scenario~1, especially for large values of~$m$. This was expected, since all the processes considered in this scenario display linear dependence structures. However, $\widehat{d}_{\text{FQA}}$ is able to reach almost perfect results when $T=600$ for small values of $m$, which is coherent with the 2DS plot in Section 4 of the Supplement. {Dissimilarities $\widehat{d}_{\text{FACF}}$ and $\widehat{d}_{\text{FSACF}}$ show substantially worse results than $\widehat{d}_{\text{FQA}}$ in all cases. In Scenario~2, the distance $\widehat{d}_{\text{FQA}}$ exhibits clearly greater clustering effectiveness than those remaining when $T=200$, which suggests that the FQA-based quantities are more appropriate for performing clustering than the competing features in situations where the time series have short to moderate lengths and the underlying groups are characterized by complex forms of functional dependence, like those produced by the fGARCH models. When $T=600$, the proposed distance again achieves very high scores in this challenging scenario associated with the 2DS plot on the right panel of Figure~1 of the Supplement. Dissimilarity $\widehat{d}_{\text{FACF}}$ also shows high clustering effectiveness in Scenario 2 when $T=600$. The results for the fuzzy C-means approaches are provided and discussed in Section 5 of the Supplement.}

Additional analyses were performed considering a different set of quantile levels for the distance $\widehat{d}_{\text{FQA}}$, and also considering functional time series of unequal lengths. The results are provided in Section 6 of the Supplement. 

\subsubsection{Scenarios with some degree of uncertainty}\label{subsubsection2as2}

Scenarios 1 and 2 assume a clustering partition formed by well-separated groups, and therefore do not include uncertain settings. To assess the impact of ambiguous time series on the clustering results, additional numerical experiments were conducted, based on two new scenarios described below.

\begin{description}
\item[Scenario 3] A collection of 11 functional time series, with five time series simulated from a FAR$(2)$ process with vector of coefficients $(c_1, c_2, c_3, c_4)$=$(-0.4, 0.5, -0.4, 0.5)$, five additional series generated from a FAR(2) process with vector of coefficients $(c_1, c_2, c_3, c_4)=(0.4, 0.5, 0.4, 0.5)$, and one isolated series generated considering an independent Brownian motion over $[0, 1]$.
\item[Scenario 4] A collection of 11 functional time series, with five time series simulated as in cluster $\mathcal{C}_2$ of Scenario 2, five additional series generated as in cluster $\mathcal{C}_3$ of Scenario 2, and one isolated series generated considering an independent Brownian motion over $[0, 1]$.
\end{description}

The series length, the number of simulation replicates, the number of discretization points, and the number of random starts, as well as the collections $\mathcal{L}$ and $\mathcal{T}$, were set in the same way as in the crisp scenarios. In these new simulations, we considered $C=2$ and we evaluated the clustering methods by means of the frequency with which: (i) the five functional time series from the first generating process were placed in a given cluster; (ii) the five functional time series from the second generating process were placed in the remaining cluster, and (iii) the membership degrees of the extra time series were moderately spread out among both groups.

The performance measure previously described requires the selection of a fixed threshold to determine when a time series must be placed in a given group. In this regard, we used a threshold of~0.7, that is, a given functional time series was assigned to a specific cluster if its membership degree in that cluster was greater than~0.7. Note that similar rules based on thresholds have already been employed for the evaluation of time series clustering methods \citep{lopez2022quantile1}.

\begin{figure}[!htb]
\centering
\includegraphics[width=0.73\textwidth]{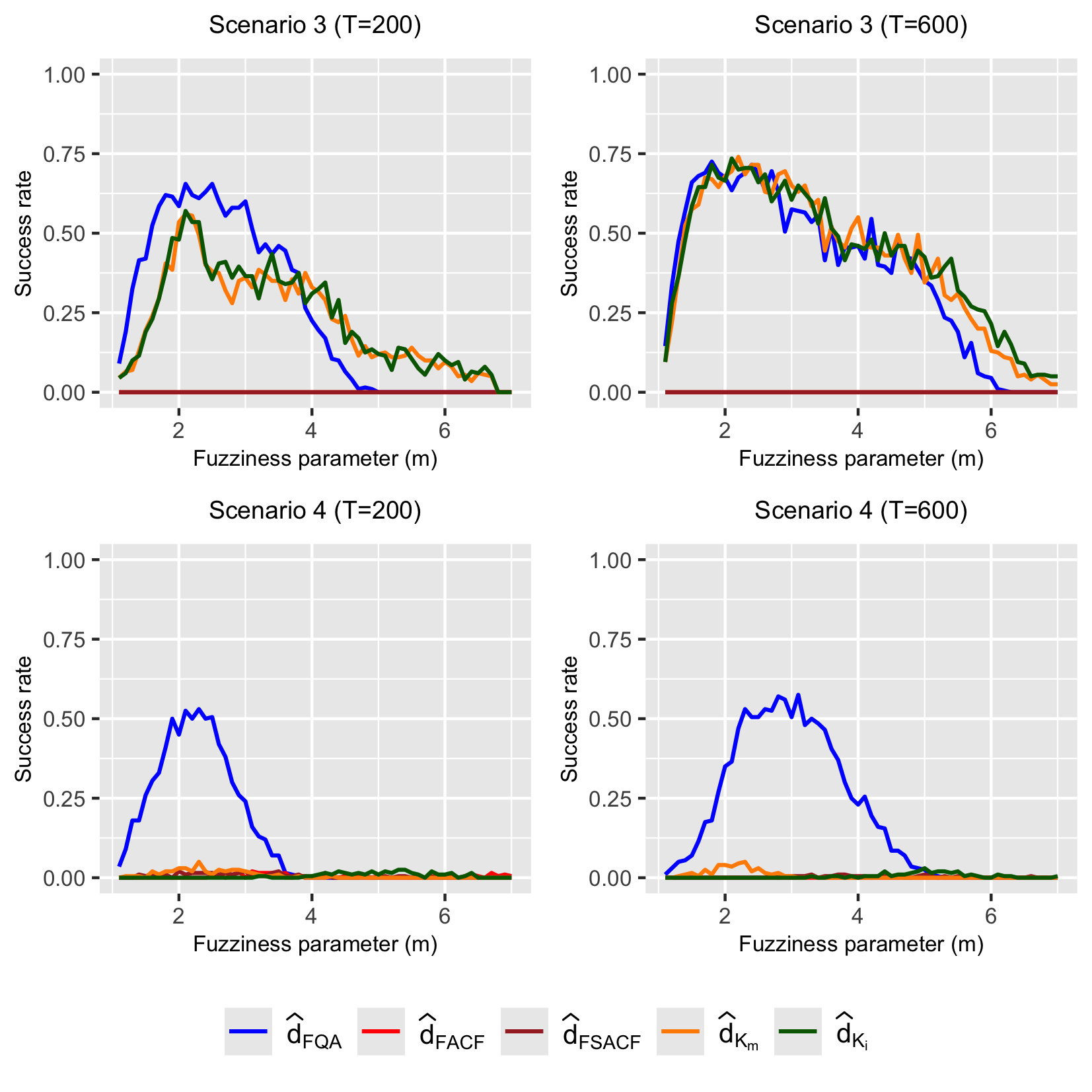}
\caption{\small Success rates with respect to $m$ for the fuzzy $C$-medoids model based on different metrics in the uncertain scenarios.}
\label{Fig_2}
\end{figure}

The assessment rule depends heavily on the value of $m$, as a failed trial may result from a time series whose membership degrees don't meet the requirements. The methods can also exhibit varying performance for different values of $m$. Therefore, each clustering procedure was executed for several fuzziness parameter values over a sufficiently large interval. The success rates for the fuzzy $C$-medoids approaches are shown in Figure~\ref{Fig_2}.

The curves in Figure~\ref{Fig_2} corroborate that the parameter $m$ substantially influences clustering accuracy. The performance for all methods is bad when very small and large values of this parameter are considered, which was expected since fuzzy solutions with all membership degrees close to 1 or 0.5 are produced in such cases. On the other hand, when $m$ is in the middle of the range, better clustering accuracy is achieved by the procedures. In Scenario 3, the proposed metric attains similar overall results compared to the dissimilarities based on Kendall autocorrelations, even though the models in this scenario are characterized by linear dependence structures. Generally, better success rates are obtained for these three distances when increasing the value of $T$. On the other hand, the metrics $\widehat{d}_{\text{FACF}}$ and $\widehat{d}_{\text{FSACF}}$ achieve success rates of 0 for both values of the series length (the red color can not be seen in the corresponding graphs because the brown and red lines are completely overlapping). In Scenario~4, the dissimilarity $\widehat{d}_{\text{FQA}}$ gets substantially better results than all the competing metrics. In particular, $\widehat{d}_{\text{FACF}}$, $\widehat{d}_{\text{FSACF}}$, $\widehat{d}_{\text{K}_\text{m}}$, and $\widehat{d}_{\text{K}_\text{i}}$ attain very poor rates of correct classification even when long time series ($T=600$) are considered. The above analyses confirm that a proper choice of the fuzziness parameter is essential. However, this topic is not addressed in this paper because several well-known mechanisms exist for the selection of this parameter; see, e.g., \cite{zhou2014fuzziness}. 

Figure 2 in the Supplement shows the rates of correct classification as a function of $m$ for the fuzzy $C$-means procedures. The results are very similar to those in Figure~\ref{Fig_2}, with the dissimilarity $\widehat{d}_{\text{FQA}}$ achieving results comparable to $\widehat{d}_{\text{K}_\text{m}}$ and $\widehat{d}_{\text{K}_\text{i}}$ in Scenario 3, and dramatically outperforming all the alternative metrics in Scenario 4.

\begin{table}[!htb]
\centering
\tabcolsep 0.4in
\caption{\small {Maximum value and area under the fuzziness curve  (in brackets) for the fuzzy $C$-medoids model based on several dissimilarities in the uncertain scenarios. The highest scores are highlighted in bold.}}\label{tabledq}
\begin{tabular}{@{}llllll@{}}
\toprule 
               & \multicolumn{2}{c}{Scenario 3} & \multicolumn{2}{c}{Scenario 4} \\
 Metric  & $T=200$  & $T=600$ & $T=200$  & $T=600$   \\\midrule
 $\widehat{d}_{\text{FQA}}$ & \textbf{0.66} (\textbf{1.51})  & 0.73 (2.29) & \textbf{0.53} (\textbf{0.75}) & \textbf{0.58} (\textbf{1.15})     \\
 $\widehat{d}_{\text{FACF}}$ & 0.00 (0.00)  & 0.00 (0.00) & 0.02 (0.02) & 0.00 (0.00)   \\
  $\widehat{d}_{\text{FSACF}}$   & 0.00 (0.00) & 0.00 (0.00) & 0.02 (0.03) & 0.01 (0.01)      \\
  $\widehat{d}_{\text{K}_\text{m}}$   & 0.56 (1.33) & \textbf{0.74} (2.50) & 0.05 (0.04) &  0.05 (0.04)      \\
  $\widehat{d}_{\text{K}_\text{i}}$   &  0.57 (1.37) & 0.73 (\textbf{2.57}) & 0.03 (0.03) &  0.03 (0.03)    \\
\bottomrule
\end{tabular}
\end{table}

Rigorous comparisons based on Figure~\ref{Fig_2} and Figure 2 in the Supplement can be made by computing 
\begin{inparaenum}
\item[(i)] the maximum value of each curve, and 
\item[(ii)] the area under each curve, denoted by the area under the fuzziness curve, which was already used by \cite{lopez2022quantile1} and \cite{lopez2023two} in the context of fuzzy clustering of time series. 
\end{inparaenum}
Table~\ref{tabledq} contains the corresponding values for the fuzzy $C$-medoids models {(the values for the fuzzy $C$-means models are given in Table~6 of the Supplement)}, which confirm the nice overall behavior of the proposed metric $\widehat{d}_{\text{FQA}}$ in Scenarios~3 and~4. {The simulation experiments described above were repeated considering different values for the threshold (namely 0.6 and 0.8), but the corresponding results were not substantially different from those presented. For this reason, we decided not to include these additional analyses.}

In summary, the experiments in Section~\ref{subsectionedr} show that clustering approaches based on $\widehat{d}_{\text{FQA}}$: 
\begin{inparaenum} 
\item[(i)] perform competitively with linear time series, even outperforming Kendall autocorrelation-based methods in the presence of ambiguous series, and 
\item[(ii)] achieve significantly higher accuracy when the underlying clusters exhibit complex forms of functional dependence. 
\end{inparaenum}
An analysis of the efficiency of the clustering approaches examined throughout Section~\ref{sectionsimulations} was also performed. The results are shown in Section 8 of the Supplement. 

\section{Financial and demographic applications}\label{sectionapplication}

We show an application of the proposed clustering approaches using high-frequency financial time series, preceded by exploratory analyses. A second application on age-specific mortality improvement rates across 41 countries is presented in Section 10 of the Supplement.

\subsection{Clustering S\&P 500 companies}\label{subsectiona1}

We use a dataset of closing prices from 40 S\&P 500 companies, recorded in 5-minute intervals between 9:30 and 16:00 EST, covering the period from 2 January 2018 to 31 December 2020. The data, sourced from Refinitiv Datascope (\href{https://select.datascope.refinitiv.com/DataScope/}{https://select.datascope.}), includes companies from two sectors: 20 from communication services (CS) and 20 from energy (EN). The names and symbols for the 40 companies are provided below.

\begin{itemize}
\item {\textbf{CS sector}. Alphabet Inc. A (GOOGL), Alphabet Inc. B (GOOG), AT\&T (T), Charter Communications (CHTR), Cisco Systems (CSCO), Comcast (CMCSA), Electronic Arts (EA), Fox Corporation. A (FOXA), Fox Corporation. B (FOX), Interpublic Group of Companies (IPG), Live Nation Entertainment (LYV), Meta Platforms (META), Netflix (NFLX), News Corporation. A (NWSA), News Corporation. B (NWS), Omnicom Group (OMC), T-Mobile US (TMUS), Take-Two Interactive (TTWO), Verizon (VZ), and Walt Disney (DIS).}
\item {\textbf{EN sector}. APA Corporation (APA), Baker Hughes (BKR), Chevron Corporation (CVX), Conoco Phillips (COP), Devon Energy (DVN), Diamondback Energy (FANG), EOG Resources (EOG), EQT Corporation (EQT), ExxonMobil (XOM), Halliburton (HAL), Hess Corporation (HES), Kinder Morgan (KMI), Marathon Oil (MRO), Marathon Petroleum (MPC), Occidental Petroleum (OXY), ONEOK (OKE), Phillips 66 (PSX), Pioneer Natural Resources (PXD), Schlumberger (SLB), and Valero Energy (VLO).}           
\end{itemize}

\begin{figure}[!htb]
\centering
\begin{subfigure}{.5\textwidth}
\centering
\includegraphics[width=0.95\linewidth]{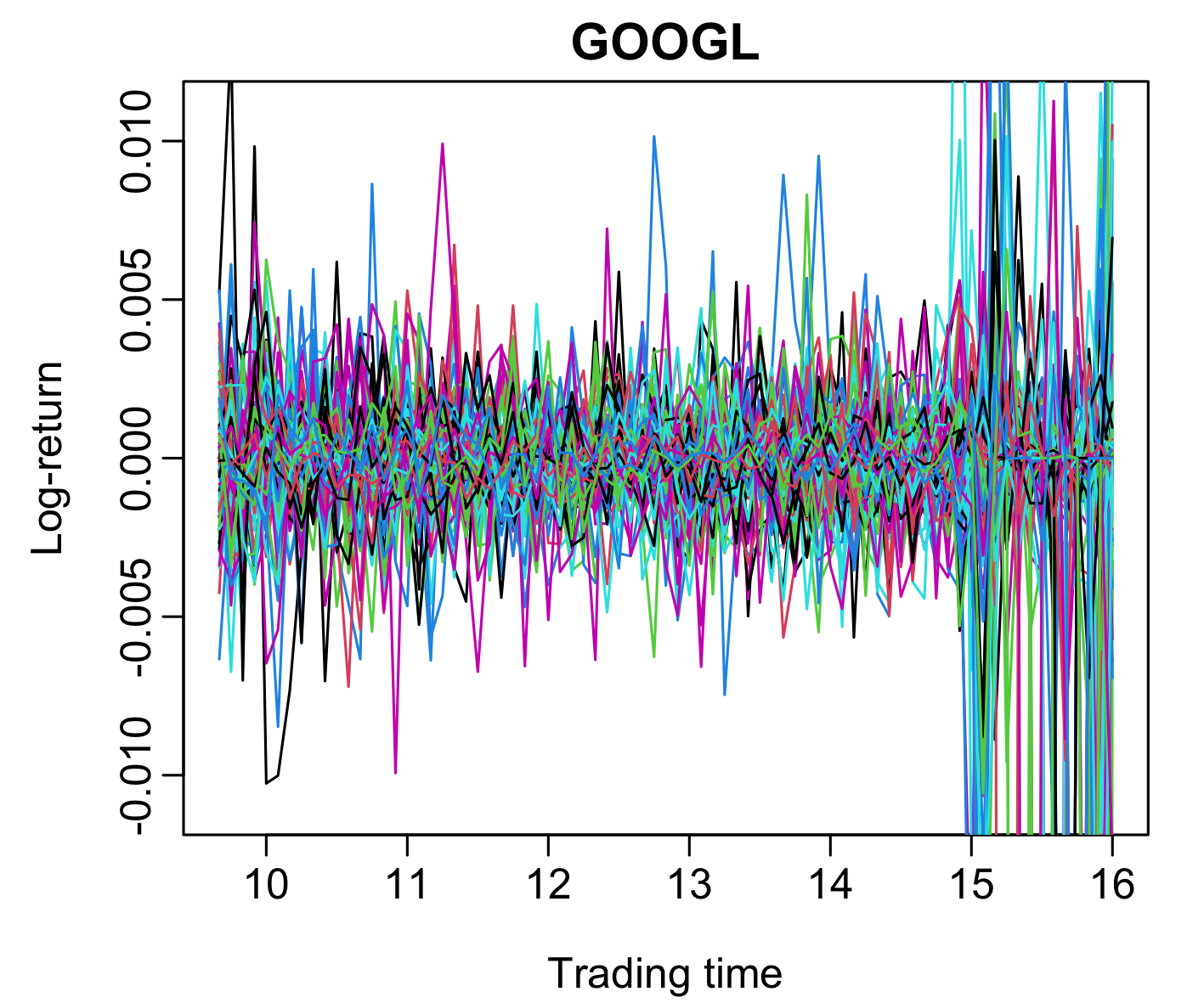}
\end{subfigure}%
\begin{subfigure}{.5\textwidth}
\centering
\includegraphics[width=0.95\linewidth]{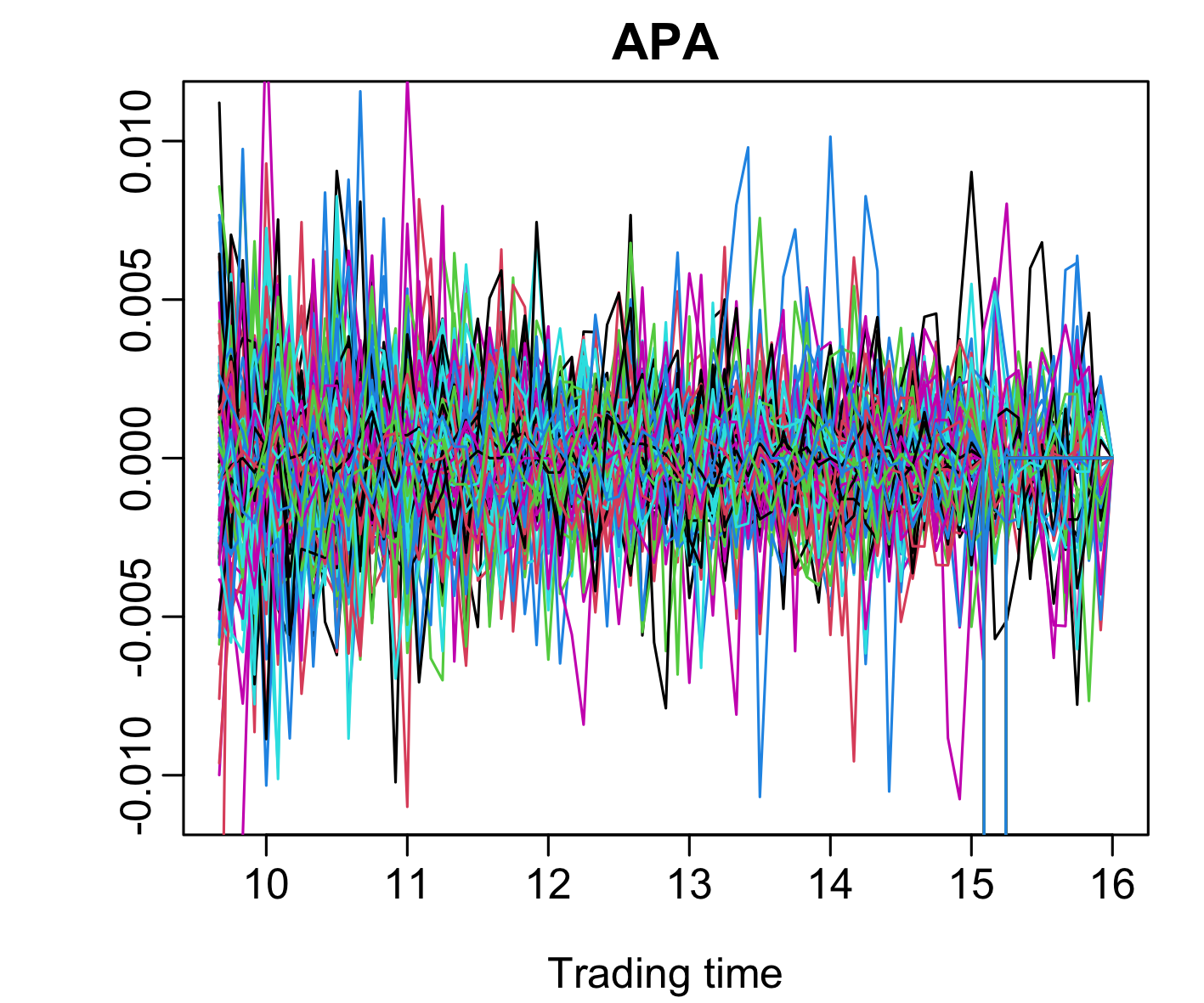}
\end{subfigure}
\caption{\small First 100 curves of the functional time series of log-returns for the companies GOOGL and APA.}
\label{Fig_4}
\end{figure}

This dataset was used to construct a collection of 40 functional time series (one for each company) by considering the daily curves of stock prices during the corresponding period. As the number of trading days in a given year is 252, and the prices are recorded in a 5-minute resolution during 6.5 hours each day, the corresponding functional time series have length $T=756$ and are observed on $p=78$ points. As the functional time series of prices are not stationary, these series are transformed into so-called intraday log returns. Denoting $P_{i,t}(u_j)$ as the intraday 5-minute closing price of the $i$\textsuperscript{th} company at time $u_j$ on trading day $t$, the corresponding sequence of log-returns is defined as 
\begin{equation*}
R_{i,t}\left(u_j\right)=\ln P_{i,t}(u_j)-\ln P_{i,t}(u_{j-1}),
\end{equation*}
where $i=1,2, \ldots, 40$, $j=2,3, \ldots, 78$ and $t=1,2, \ldots, 756$. In Figure~\ref{Fig_4}, the first 100 curves (corresponding to 2018) of the functional time series of log returns for the companies GOOGL and APA are displayed. Although the values of the $x$-axis in these plots were chosen according to the daily trading times, the functional time series of log returns are assumed to be defined in 77 evenly spaced points on the interval $[0,1]$.

Stock returns exhibit complex dependence beyond linearity. Given the proposed clustering algorithms' superior performance with time series displaying complex dependence patterns (see Scenarios 2 and 4 in Section~\ref{subsectionedr}), using functional time series of log-returns is a reasonable choice for applying the procedure.

From a financial perspective, the temporal behavior of log-return curves is likely influenced by the sector of the company. Analyses in \cite{lopez2021quantile} suggest that log-returns from certain S\&P 500 sectors exhibit similar dynamics. Additionally, we expect to identify ambiguous time series using a fuzzy approach.

The proposed fuzzy $C$-medoids and fuzzy $C$-means models based on $\widehat{d}_{\text{FQA}}$ were applied to the dataset of functional time series of log-returns considering~$C=2$ to determine to what extent the algorithms can differentiate between the time series associated with both sectors.  The remaining parameters (see Section~\ref{subsectionhs}) were set as $\mathcal{T}=\{0.1, 0.5, 0.9\}$, $\mathcal{L}=\{1\}$ and $m=1.8$. Concerning $\mathcal{L}$ and $m$, the considered values for the selection process were $\mathcal{L}\in\big\{\{1\}, \{1, 2\}, \{1, 2, \ldots, 10\}\big\}$ and $m \in \{1.1, 1.2, \ldots, 2\}$. {We should note that, regarding the selection of parameter $m$, we employed the procedure introduced in Section \ref{subsubsectionhs3} for the fuzzy $C$-means model (since the Xie-Beni index is originally defined in the context of fuzzy $C$-means) and then used the resulting value for $m$ (1.8) also for the fuzzy $C$-medoids model. In this way, a direct comparison between the membership degrees of the corresponding fuzzy partitions can be carried out.}

As an exploratory exercise, we constructed a 2DS plot based on $\widehat{d}_{\text{FQA}}$, which was computed considering the mentioned values for $\mathcal{T}$ and $\mathcal{L}$. The results are given in Section 9 of the Supplement. The fuzzy partitions returned by the $\widehat{d}_{\text{FQA}}$-based models are shown in Table~\ref{tablefuzzysectors}. Superscripts~\textsuperscript{1} and~\textsuperscript{2} in the name of the companies indicate the medoids associated with clusters $\mathcal{C}_1$ and $\mathcal{C}_2$, respectively, for the fuzzy $C$-medoids model. Membership degrees greater than 0.7 are highlighted in bold.
\begin{table}[!htb]
\centering
\caption{\small Fuzzy solution ($C=2$) produced by the fuzzy $C$-medoids and the fuzzy $C$-means procedures based on $\widehat{d}_{\text{FQA}}$ when grouping the 40 functional time series of log-returns. The superscripts 1 and 2 indicate the medoids for clusters $\mathcal{C}_1$ and $\mathcal{C}_2$, respectively. Membership degrees above 0.7 are written in bold.}\label{tablefuzzysectors}
\tabcolsep 0.065in
\begin{tabular}{@{}lllll|lllll@{}} 
\toprule
Company       & \multicolumn{2}{c}{$\mathcal{C}_1$} & \multicolumn{2}{c}{$\mathcal{C}_2$}    & Company    &   \multicolumn{2}{c}{$\mathcal{C}_1$} & \multicolumn{2}{c}{$\mathcal{C}_2$}   \\
(CS sector)   & medoids  & means & medoids  & means   & (EN sector)   & medoids  & means & medoids  & means \\\midrule
GOOGL &       0.58 & {0.70} & 0.42 & {0.30} &   APA  & 0.09 & {0.06} & \textBF{0.91} & {\textbf{0.94}}       \\
GOOG &  0.59 & {0.69} & 0.41 & {0.31} &       BKR  &  0.57 & {\textbf{0.78}} & 0.43 & {0.22}   \\
T &     0.63 &  {0.41} & 0.37 & {0.59} &  CVX      & 0.19 & {0.08} & \textBF{0.81} & {\textbf{0.92}}     \\
CHTR  & \textBF{0.73} & {\textbf{0.98}} & 0.27 & {0.02} &       COP  & 0.11 & {0.03} & \textBF{0.89} & {\textbf{0.97}}    \\
CSCO &  \textBF{0.87} & {\textbf{0.85}} &  0.13 & {0.15} &        DVN  & 0.12 & {0.02} & \textBF{0.88} & {\textbf{0.98}}   \\
    CMCSA &  \textBF{0.78} & {\textbf{0.94}} &  0.22 & {0.06} &  FANG     & 0.18 & {0.04} & \textBF{0.82} & {\textbf{0.96}}      \\
EA &  \textBF{0.71} & {\textbf{0.93}} &   0.29 & {0.07} &     EOG    & 0.16 & {0.05} & \textBF{0.84} & {\textbf{0.95}}       \\
FOXA &  \textBF{0.80} & {\textbf{0.96}} & 0.20 & {0.04} &     EQT   & 0.31 & {0.13} & 0.69 & {\textbf{0.87}}     \\
FOX &  \textBF{0.74} & {\textbf{0.97}} & 0.26 & {0.03} &        XOM   & 0.17 & {0.06} & \textBF{0.83} & {\textbf{0.94}}     \\
IPG$^1$ &  \textBF{1.00} & {\textbf{0.74}} & 0.00 & {0.26} &     HAL  & 0.27 & {0.38} & \textBF{0.73} & {0.62}   \\
LYV &  0.15 & {0.11} &  \textBF{0.85} & {\textbf{0.89}} &        HES    & 0.43 & {0.29} & 0.57 & {\textbf{0.71}}   \\
META &   \textBF{0.76} & {\textbf{0.90}} & 0.24 & {0.10} &       KMI  & 0.15 & {0.08} & \textBF{0.85} & {\textbf{0.92}}   \\
NFLX  &  \textBF{0.76} & {\textbf{0.98}} &  0.24 & {0.02} &      MRO$^2$ & 0.00 & {0.03} & \textBF{1.00} & {\textbf{0.97}}     \\
NWSA & \textBF{0.90} & {\textbf{0.87}} &  0.10 & {0.13} &       MPC   & 0.11 & {0.03} & \textBF{0.89} & {\textbf{0.97}}   \\
NWS &  \textBF{0.86} & {\textbf{0.90}} &  0.14 & {0.10} &    OXY    & 0.20 & {0.07} & \textBF{0.80} & {\textbf{0.93}}  \\
OMC  & 0.05 & {0.03} & \textBF{0.95} & {\textbf{0.97}} &     OKE   & 0.25 & {0.15} & \textBF{0.75} & {\textbf{0.85}}      \\
TMUS    &  \textBF{0.73} & {\textbf{0.95}} & 0.27 & {0.05} &     PSX  & 0.18 & {0.11} & \textBF{0.82} & {\textbf{0.89}}    \\
TTWO   & 0.68 & {\textbf{0.88}} & 0.32 & {0.12} & PXD      & 0.29 & {0.15} & \textBF{0.71} & {\textbf{0.85}} \\
VZ   & \textBF{0.74} & {0.52} & 0.26 & {0.48} &    SLB    & 0.26 & {0.13} & \textBF{0.74} & {\textbf{0.87}}       \\
DIS  & \textBF{0.82} & {0.52} &  0.18 & {0.48} &     VLO    & 0.17 & {0.06} & \textBF{0.83} & {\textbf{0.94}} \\ \bottomrule
\end{tabular}
\end{table}

For both clustering models, the membership degrees in Table~\ref{tablefuzzysectors} are coherent with the 2DS plot in Figure~3 of the Supplement. Cluster $\mathcal{C}_1$ contains most of the companies of the CS sector, while the opposite occurs for cluster $\mathcal{C}_2$ with the EN sector companies. In fact, more than half of the series in each group exhibit a maximum membership degree greater than 0.70 (and often rather close to 1 in the case of the fuzzy $C$-means model), which suggests that the latent clustering configuration is formed by moderately well-defined groups. However, there are three series showing rather ambiguous behavior (i.e., with both memberships no greater than 0.7) for both clustering approaches, which correspond to the CS companies GOOGL, GOOG, and T. These functional time series share serial dependence patterns characterizing both groups, and the corresponding companies could be carefully analyzed to understand why their log returns show such ambiguous behavior. Note that this type of analysis is possible thanks to the flexibility that the fuzzy paradigm provides through the membership degrees. From the 2DS plot in Figure~3 of the Supplement, there are two CS companies showing a high membership degree in cluster $\mathcal{C}_2$, namely, LYV and OMC (the corresponding time series are associated with the two rightmost red points). This could be caused by the occurrence of abnormal financial events in these companies during the analyzed period. In any case, confirming such a hypothesis is beyond the scope of this paper. 

The fuzzy solutions associated with the fuzzy $C$-medoids and the fuzzy $C$-means procedures based on $\widehat{d}_{\text{FACF}}$, $\widehat{d}_{\text{FSACF}}$, $\widehat{d}_{\text{K}_\text{m}}$, and $\widehat{d}_{\text{K}_\text{i}}$ considering $C=2$ were also obtained. The selection of $m$ was carried out similarly as in the above analysis, and the corresponding fuzzy partitions were obtained for the selected values for $m$, namely $m=2$ for the four alternative metrics. The values of ARIF and JIF associated with the two fuzzy procedures based on $\widehat{d}_{\text{FQA}}$ and the remaining dissimilarities were obtained (we assumed that the true partition is given by the considered sectors, CS and EN). Note that these values indicate the ability of each distance to differentiate between the underlying sectors. However, such values are clearly influenced by the choice of $m$, which is different among metrics. In order to circumvent this issue, we also computed the classical versions of the Adjusted Rand Index (ARI) and the Jaccard Index (JI) by transforming the corresponding fuzzy partitions into crisp partitions via the maximum membership rule. 

The results are displayed in Table~\ref{indexesapplication}. The procedures based on the proposed metric achieve acceptable values for the fuzzy clustering indexes (ARIF and JIF), which was expected given the fuzzy partition in Table~\ref{tablefuzzysectors}. The metrics $\widehat{d}_{\text{FACF}}$, $\widehat{d}_{\text{FSACF}}$, and $\widehat{d}_{\text{K}_\text{i}}$ attain better scores than $\widehat{d}_{\text{FQA}}$ for both indexes when the fuzzy $C$-medoids models are considered. However, for both clustering approaches, $\widehat{d}_{\text{FQA}}$ reaches substantially higher values than all the remaining dissimilarities in terms of the crisp clustering indexes (ARI and JI). These results corroborate that $\widehat{d}_{\text{FQA}}$ is capable of detecting some structure in the functional time series dataset having an actual meaning in terms of industrial sectors. 

\begin{table}[!htb]
\centering
\tabcolsep 0.4in
\caption{\small {Clustering accuracy for the fuzzy $C$-medoids and the fuzzy $C$-means (in brackets) procedures based on different metrics when grouping the 40 functional time series of log-returns. The true partition is determined by the corresponding sectors (CS and EN). The highest scores are highlighted in bold.}}\label{indexesapplication}
{\begin{tabular}{@{}lcccc@{}} 
\toprule
Metric   & ARIF & JIF & ARI & JI \\\midrule
$\widehat{d}_{\text{FQA}}$ ($m=1.8$)  &  0.33 (\textbf{0.40})                    &        0.49 (\textbf{0.54}) & \textBF{0.72} (\textbf{0.63}) & \textBF{0.75} (\textbf{0.68})                    \\
{$\widehat{d}_{\text{FACF}}$ ($m=2$)} &  {\textbf{0.40}} (\textbf{0.40})                    &        {\textbf{0.54}} (\textbf{0.54}) & {0.41} (0.41) & {0.55} (0.55)                  \\
{$\widehat{d}_{\text{FSACF}}$ ($m=2$)} &  {0.37} (0.37)                 &        {0.52} (0.52) & {0.41} (0.41) & {0.55} (0.55)                    \\
$\widehat{d}_{\text{K}_\text{m}}$ ($m=2$)    & 0.23 (0.24)                     &   0.46 (0.46) &  0.23 (0.23) & 0.47 (0.47)                           \\
$\widehat{d}_{\text{K}_\text{i}}$ ($m=2$)  & 0.37 (0.37)                        &   0.53 (0.53)   & 0.41 (0.41) & 0.55 (0.55) \\ \bottomrule
\end{tabular}}
\end{table}

When performing feature-based fuzzy clustering of time series, it is common to summarize the characteristics of the elements in each group by considering a weighted average of the considered features, with the weights given by the corresponding membership degrees. For a given cluster $c \in \{1, 2\}$, the dependence structures characterizing that group can be described through the set of features $\big\{\overline{\rho}_c(\tau_1, \tau_2, 1, \tau_1, \tau_2): \tau_1, \tau_2 \in \{0.1, 0.5, 0.9\}\big\}$, with
\begin{equation}\label{wa}
\overline{\rho}_c(\tau_1, \tau_2, 1, \tau_1, \tau_2)=\frac{\sum_{i=1}^{40}u_{ic}^*\widehat{\rho}^{(i)}(\tau_1, \tau_2, 1, \tau_1, \tau_2)}{\sum_{i=1}^{40}u_{ic}^*},
\end{equation}
where $u_{ic}^{*}$ denotes the resulting membership degree of the $i$\textsuperscript{th} functional time series in the $c$\textsuperscript{th} cluster according to the fuzzy partition produced by one of the $\widehat{d}_{\text{FQA}}$-based clustering algorithms (see Table~\ref{tablefuzzysectors}).

The aforementioned quantities are given in Table~\ref{wavg} for the fuzzy $C$-medoids model. All the considered values are similar for both clusters, which indicates that the 40 functional time series of log-returns exhibit a quite homogeneous dynamic behavior, thus highlighting the complexity of the clustering task. However, slight-to-moderate differences can be observed for some of the quantities. In particular, the highest discrepancies occur when $(\tau_1, \tau_2) \in \big\{(0.1, 0.1), (0.1, 0.9), (0.9, 0.1)\big\}$. Note that all the features associated with $\mathcal{C}_2$ are larger, in absolute value, than those associated with $\mathcal{C}_1$, which suggests that the functional time series in the second group (EN sector) display a stronger degree of serial dependence. In addition, although interpreting these quantities is not straightforward, a careful analysis of the corresponding values could give the practitioner a general picture of the different dynamics displayed by the companies in both sectors.
\begin{table}[!htb]
\centering
\tabcolsep 0.245in
\caption{\small Values of $\overline{\rho}_c(\tau_1, \tau_2, 1, \tau_1, \tau_2)$ for the two-cluster solution produced by the fuzzy $C$-medoids procedure based on $\widehat{d}_{\text{FQA}}$ with the 40 functional time series of log-returns.}\label{wavg}
\begin{tabular}{@{}lrrrlrrrr@{}} 
\toprule
                & \multicolumn{3}{c}{$\tau_2$} & & & \multicolumn{3}{c}{$\tau_2$} \\
                \cmidrule{2-5}\cmidrule{7-9}
$\mathcal{C}_1$ & 0.1   & 0.5   & 0.9   & & $\mathcal{C}_2$  & 0.1   & 0.5   & 0.9  \\ \midrule 
$\tau_1=0.1$    & 0.51  & -0.08 & -0.54 & & $\tau_1=0.1$     & 0.60  & -0.09 & -0.62 \\
$\tau_1=0.5$    & -0.07 & 0.08  &  0.07 & & $\tau_1=0.5$     & -0.08 & 0.13  & 0.09  \\
$\tau_1=0.9$    & -0.50 & 0.09  & 0.53  & & $\tau_1=0.9$     & -0.58 & 0.10  & 0.59  \\
\bottomrule
\end{tabular}
\end{table}

\section{Conclusions}\label{sectionconclusions}

We introduced a distance measure for functional time series based on functional quantile autocorrelation, an extension of the quantile autocorrelation for real-valued time series. This dissimilarity was used with fuzzy $C$-medoids and fuzzy $C$-means algorithms to design clustering techniques for functional time series. These methods create a soft partition, allowing time series to have varying membership degrees in clusters, handling uncertainty in temporal datasets. The techniques involve hyperparameters, which can be selected using reasonable heuristic rules. 
	
The proposed clustering procedures were evaluated through simulations with well-defined and fuzzy group settings, using various functional processes. We compared them with four clustering algorithms based on functional autocorrelations and one using functional panel data modeling. The proposed techniques:
\begin{inparaenum}
\item[(i)] yield competitive results on linear time series, though they are generally outperformed by functional Kendall autocorrelation-based methods in this context, and  
\item[(ii)] significantly outperform alternative approaches when the clusters exhibit complex functional dependence, while also being computationally efficient.
\end{inparaenum} 
We recommend the proposed methods for predominantly nonlinear time series and functional Kendall autocorrelation-based methods for mostly linear series.

Two applications demonstrated the proposed methods' effectiveness, showing that:
\begin{inparaenum}
\item[(i)] the proposed measure of serial dependence reveals key insights into the nature of the analyzed time series, and  
\item[(ii)] examining the membership degrees in the clustering solutions offers practitioners a deeper understanding of the underlying structure of the time series.
\end{inparaenum}

This work can be extended in several ways:  
\begin{inparaenum}
\item[1)] Improving hyperparameter selection for greater accuracy and efficiency.  
\item[2)] Developing robust extensions using metric, noise, and trimmed approaches \citep{lopez2022quantile2}.
\item[3)] Incorporating spatial components to handle spatially structured functional time series \citep{lopez2022spatial}.  
\item[4)] Defining a spectral counterpart of functional quantile autocorrelation for frequency analysis.  
\item[5)] Studying its asymptotic behavior to construct statistical tests.
\end{inparaenum}

\bigskip
\begin{center}
{\large\bf Supplement}
\end{center}

\begin{description}
\item[\Rlogo\ code for clustering high-dimensional functional time series] The \Rlogo\ code for Monte-Carlo simulation studies and empirical application is available at \url{https://github.com/anloor7/PostDoc/tree/main/r_code/functional}.
\item[Supplement file] {The supplement file includes some additional contents related to the proposed clustering algorithms and the corresponding hyperparameter selection problem (Sections~1\textendash3), the 2DS representations associated with Scenarios 1 and 2 in Section~\ref{sectionsimulations} (Section~4), some simulation results for the proposed models (Sections~5\textendash7), an evaluation of the time complexity of the proposed and the competing approaches (Section~8), the 2DS representations associated with the financial application in Section~\ref{subsectiona1} (Section~9), and the results for the demographic application (Section~10).}
\end{description}

\section*{Acknowledgments}

The authors sincerely thank the two referees for their valuable feedback, which greatly improved the paper. The first and second authors acknowledge the support of King Abdullah University of Science and Technology (KAUST), while the third author thanks the Australian Research Council for funding through Discovery Project DP230102250 and Future Fellowship FT240100338.

\newpage

\begin{center}
\title{\bf \large Supplement to ``Dependence-based fuzzy clustering of functional time series"}
\end{center}

\spacingset{1.4}

\renewcommand{\thesection}{S\arabic{section}}
\setcounter{section}{0}

\section{Algorithm for the fuzzy \texorpdfstring{$C$}{C}-medoids method}

{Algorithm \ref{algorithm1} contains an outline of the fuzzy $C$-medoids model based on $\widehat{d}_{\text{FQA}}$.}

\begin{algorithm}
\textcolor{black}{\caption{\textcolor{black}{Fuzzy $C$-medoids algorithm based on the metric $\widehat{d}_{\text{FQA}}$. \label{algorithm1}}}}
\begin{algorithmic}
\State \textcolor{black}{Fix $C$, $m$ and \textit{max.iter}} 
\State \textcolor{black}{Set $iter \, =0$}
\State \textcolor{black}{Select the initial medoids, $\widetilde{\mathbb{S}}=\{\widetilde{\boldsymbol{\mathcal{X}}}^{(1)}, \ldots, \widetilde{\boldsymbol{\mathcal{X}}}^{(C)}\}$}
\Repeat
\State \textcolor{black}{Fix $\widetilde{\mathbb{S}}_{\text{OLD}}=\widetilde{\mathbb{S}}$}
\Comment{\textcolor{black}{Record the current medoids}}
\State \textcolor{black}{Obtain $u_{ic}$, $i=1,\ldots,n$, $c=1,\ldots,C$, using~(3) in the main text}
\State \textcolor{black}{For each cluster $c \in \{1,\ldots,C\}$, compute the index $j_c \in \{1,\ldots,n\}$ as indicated in Section 3.1 in the main text}
\State \textbf{return} \textcolor{black}{$\widetilde{\boldsymbol{\mathcal{X}}}^{(c)}=\boldsymbol{\mathcal{X}}^{(j_c)}$, for $c=1,\ldots,C$}  
\Comment{\textcolor{black}{Update the medoids}}
\State \textcolor{black}{$iter =  iter \, + 1$}
\Until{ \mbox{ \textcolor{black}{$\widetilde{\mathbb{S}}_{\text{OLD}}=\widetilde{\mathbb{S}} \mbox{ or } iter \, = \, max.iter$}} } 
\State \textbf{return} \textcolor{black}{The final clustering solution and the corresponding collection of medoids}
\end{algorithmic}
\end{algorithm}

\newpage
\section{Using \texorpdfstring{$\widehat{d}_{\text{FQA}}$}{} in combination with the fuzzy \texorpdfstring{$C$}{C}-means model}\label{sectionfcmeans}

One well-known alternative to the fuzzy $C$-medoids clustering model introduced above is the fuzzy $C$-means model \citep{bezdek1984fcm}. Following the context of Section 3.1 in the main text, denote by $\boldsymbol{\widehat{\rho}}=\left\{\boldsymbol{\widehat{\rho}}^{(1)}, \ldots, \boldsymbol{\widehat{\rho}}^{(n)}\right\}$ the set of feature vectors associated with the collection $\mathbb{S}$. Generalizing to the $n$ time series the notation used in Section 2.2 in the main text, we will assume that each vector $\boldsymbol{\widehat{\rho}}^{(i)}$ is constructed by concatenating together all the features of the form $\frac{1}{\sqrt{4LP^2B^2}}\widehat{\rho}^{(i)}(\tau_{i_1}, \tau_{i_2}, l_k, \beta_{j_1}, \beta_{j_2})$, $i=1,\ldots,n$. In this way, each vector $\boldsymbol{\widehat{\rho}}^{(i)}$ is formed by $LP^2B^2$ FQA-based features, which are multiplied by the constant $\frac{1}{\sqrt{4LP^2B^2}}$ so that the squared Euclidean distance between two vectors in $\boldsymbol{\widehat{\rho}}$ fulfills the definition of $\widehat{d}_{\text{FQA}}$. The fuzzy $C$-medoids model based on the dissimilarity $\widehat{d}_{\text{FQA}}$ looks for the set of centroids, $\boldsymbol{\overline{\widehat{\rho}}}=\left\{\boldsymbol{\overline{\widehat{\rho}}}^{(1)}, \ldots, \boldsymbol{\overline{\widehat{\rho}}}^{(C)}\right\}$, and the matrix $\boldsymbol U$ of membership degrees defining the solution of the minimization problem
\begin{equation*}
\min_{\boldsymbol{\overline{\widehat{\rho}}}, \boldsymbol U}\sum_{i=1}^{n}\sum_{c=1}^{C}u_{ic}^m\times \left\|\widehat{\boldsymbol{\rho}}^{(i)}-\overline{\widehat{\boldsymbol{\rho}}}^{(c)}\right\|^2,  \quad \text{ with respect to} \quad \sum_{c=1}^{C}u_{ic}=1, \ u_{ic} \ge 0,
\end{equation*}
where $\overline{\widehat{\boldsymbol{\rho}}}^{(c)}$ denotes the FQA-based feature vector with regards to the centroid of cluster $c$, and $u_{ic}$ and $m$ are the same as in (2) in the main text. Some extensions of the fuzzy $C$-means model previously introduced are provided by intuitionistic fuzzy $C$-means \citep{xu2010intuitionistic}, fuzzy $C$-means for big data \citep{havens2012fuzzy}, and fuzzy $C$-means based on multivariate memberships \citep{pimentel2013multivariate}.

{The previous constrained optimization problem can be solved by means of the Lagrangian multipliers method, giving rise to a two-step iterative process \citep{hoppner1999fuzzy}. One of the steps consists of the optimization of the objective function with respect to $\boldsymbol{\overline{\widehat{\rho}}}$, being $u_{ic}$ fixed:
\begin{equation}\label{updatecentroids}
\overline{\widehat{\boldsymbol{\rho}}}^{(c)}=\frac{\sum_{i=1}^n u_{i c}^m \widehat{\boldsymbol{\rho}}^{(i)}}{\sum_{i=1}^n u_{i c}^m},
\end{equation}
for $c=1, \ldots, C$.}

{The remaining step is based on the minimization of the objective function regarding $u_{ic}$, being $\boldsymbol{\overline{\widehat{\rho}}}$ fixed:
\begin{equation}\label{updatemdmeans}
u_{ic}=\left[\sum_{c^{\prime}=1}^C\left(\frac{\left\|\widehat{\boldsymbol{\rho}}^{(i)}-\overline{\widehat{\boldsymbol{\rho}}}^{(c)}\right\|^2}{\left\|\widehat{\boldsymbol{\rho}}^{(i)}-\overline{\widehat{\boldsymbol{\rho}}}^{(c')}\right\|^2}\right)^{\frac{1}{m-1}}\right]^{-1},
\end{equation}
for $i=1, \ldots, n$ and $c=1, \ldots, C$. The above two-step mechanism is repeated until the membership matrices related to two consecutive iterations are close enough or a maximum number of iterations is achieved. An outline of the fuzzy $C$-means procedure is given in Algorithm~\ref{algorithm2}.}
\begin{algorithm}
\textcolor{black}{\caption{\textcolor{black}{{Fuzzy $C$-means algorithm based on the metric $\widehat{d}_{\text{FQA}}$.} \label{algorithm2}}}}
\begin{algorithmic}
\color{black}
\State \textcolor{black}{Fix $C$, $m$, \textit{max.iter}, \textit{tol} and a matrix norm $\|\cdot\|_M$} 
\State \textcolor{black}{Set $iter \, =0$}
\State \textcolor{black}{Initialize the membership matrix, $\boldsymbol{U}=\boldsymbol{U}^{(0)}$}
\Repeat
\State \textcolor{black}{Set $\boldsymbol{U}_{\mathrm{OLD}}=\boldsymbol{U}$}
\Comment{\textcolor{black}{Record the current membership matrix}}
\State \textcolor{black}{Obtain the centroids, $\boldsymbol{\overline{\widehat{\rho}}}^{(c)}$, $c=1,\ldots,C$, using~\eqref{updatecentroids}}
\State \textcolor{black}{Compute $u_{ic}$, $i=1,\ldots,n$, $c=1,\ldots,C$, using~\eqref{updatemdmeans}}
\Comment{\textcolor{black}{Update the membership matrix}}
\State \textcolor{black}{$iter =  iter \, + 1$}
\Until{ \mbox{ \textcolor{black}{$\left\|\boldsymbol{U}-\boldsymbol{U}_{\mathrm{OLD}}\right\|_M<tol \mbox{ or } iter \, = \, max.iter$}} } 
\State \textbf{return} \textcolor{black}{The final clustering solution and the corresponding collection of centroids}
\end{algorithmic}
\end{algorithm}

\clearpage
\section{Selection of hyperparameters in the fuzzy clustering models based on \texorpdfstring{$\widehat{d}_{\text{FQA}}$}{dFQA}}

{Selection of the five hyperparameters ($C$, $m$, $\mathcal{L}$, $\mathcal{T}$ and $\mathcal{B}$) involved in the fuzzy clustering procedures based on $\widehat{d}_{\text{FQA}}$ is addressed.}

\subsection{{Selection of the set \texorpdfstring{$\mathcal{L}$}{L}}}\label{subsubsectionhs1s}

{The set of lags ($\mathcal{L}$) can be chosen by employing the selection procedure proposed by \cite{lopez2023hard} (see Section 3.4), adapted to the functional setting. This mechanism uses a criterion that consists of determining the significant lags for each functional time series in the collection and then fixing a maximum lag for all of them. In particular, given the set $\mathbb{S}=\{\boldsymbol{\mathcal{X}}^{(1)}, \boldsymbol{\mathcal{X}}^{(2)}, \ldots, \boldsymbol{\mathcal{X}}^{(n)}\}$, we suggest to select the collection $\mathcal{L}$ through the following steps:}
{\begin{enumerate}
		\item[1.] Fix $\alpha>0$ and a maximum lag $L_{\text{Max}}\in\mathbb{N}$. Using the Bonferroni\textquotesingle s adjustment for multiple comparisons, compute the corrected significance level $\alpha'=\alpha/(nL_{\text{Max}}).$ 
		\item[2.] For each series $\boldsymbol{\mathcal{X}}^{(i)} \in \mathbb{S}$:
		\begin{enumerate}
			\item[2.1.] Determine the set $\mathcal{L}_i$ formed for all the lags $l_i \in \{1, 2, \ldots, L_{\text{Max}}\}$ for which the null hypothesis of functional serial independence is rejected according to a $t$-test based on the so-called distance correlation \citep{szekely2013distance}.
			\item[2.2.] Select the lag $L_i\in \mathcal{L}_i$ associated with the lowest $p$-value.
		\end{enumerate}
		\item[3.] Set $L^*=\max\{L_1, \ldots, L_n\}$ and $\mathcal{L}=\{1, 2, \ldots, L^*\}$.  
\end{enumerate}}

{In the above process, the Bonferroni correction is used to address the problem of multiple comparisons, since $nL_{\text{Max}}$ tests are simultaneously carried out. In Step 2.2, the most significant lag is selected to avoid including redundant features, since dependence at lower lags often leads to (less strong) dependence at higher lags. In Step 3, $\mathcal{L}$ is constructed considering all lags between 1 and $L^*$. Note that $L^*$ is necessarily a significant lag for one or several functional time series, although indeed, some series may not display significant serial dependence at $L^*$ or lower lags. However, this is not an issue because the corresponding estimated features are expected to be close to zero for these series. As an example, according to the selection mechanism described above, if some of the functional time series in the collection exhibit primarily weekly dependence, then $L_i=7$ for some $i$, and the set $\mathcal{L}$ will necessarily include lag 7. The $t$-test based on the distance correlation is implemented via the function \textit{dcor.xy()} of the \Rlogo\ package \textbf{fda.usc} \citep{fda.usc}.} 

\subsection{{Selection of the sets \texorpdfstring{$\mathcal{T}$}{T} and \texorpdfstring{$\mathcal{B}$}{B}}}\label{subsubsectionhs2}

{For the collection of quantile levels ($\mathcal{T}$), we suggest using three quantiles of levels 0.1, 0.5, and 0.9, since several works on quantile-based time series clustering \citep{lafuente2016clustering,vilar2018quantile, lopez2021quantile, lopez2022quantile1} have shown that: 
\begin{inparaenum}
    \item(i) such choice often leads to a high clustering effectiveness, and 
    \item[(ii)] considering more quantile levels usually results in nonsignificant improvements. 
\end{inparaenum}    
In fact, the numerical experiments performed through Section~4 in the main text (in which different choices for $\mathcal{T}$ are considered) corroborate that $\mathcal{T}=\{0.1, 0.5, 0.9\}$ is also a reasonable choice in the functional setting. With respect to the set of thresholds ($\mathcal{B}$), we suggest to consider $\tau_{i_1}=\beta_{j_1}$ and $\tau_{i_2}=\beta_{j_2}$ (see Remark~1 in the main text) in the framework of Section 2.2 in the main text. In other words, we propose to consider the following alternative version for the distance $\widehat{d}_{\text{FQA}}$:} 
{\begin{equation*}
\widehat{d}_{\text{FQA}}^*\big(\mathcal{X}_t^{(1)}, \mathcal{X}_t^{(2)}\big)= \frac{1}{4LP^2}\sum_{k=1}^{L}\sum_{i_1=1}^{P}\sum_{i_2=1}^{P}\Big(\widehat{\rho}^{(1)}(\tau_{i_1}, \tau_{i_2}, l_k, \tau_{i_1}, \tau_{i_2})-\widehat{\rho}^{(2)}(\tau_{i_1}, \tau_{i_2}, l_k, \tau_{i_1}, \tau_{i_2})\Big)^2.
\end{equation*}}

It is worth noting that several analyses conducted in the context of Section~4 in the main text (which are not provided in the manuscript for the sake of simplicity) demonstrated that the clustering accuracy achieved using $\widehat{d}_{\text{FQA}}^*$ is often comparable to that obtained with the more general version of the distance (which includes a larger number of FQA-based terms in the summation). For this reason, the reduced version of the dissimilarity ($\widehat{d}_{\text{FQA}}^*$) will be used moving forward, although we will continue to use the notation $\widehat{d}_{\text{FQA}}$ for convenience.

\subsection{{Selection of \texorpdfstring{$C$}{C} and \texorpdfstring{$m$}{m}}}\label{subsubsectionhs3}

{For fixed $\mathcal{L}$, $\mathcal{T}$ and $\mathcal{B}$, the number of clusters ($C$) and the fuzziness parameter ($m$) can be selected by executing the corresponding clustering procedure for several choices of $(C, m)$ and picking the one minimizing the \citeauthor{xie1991validity}'s \citeyearpar{xie1991validity} index. Using the notation employed in Section \ref{sectionfcmeans}, for a given $\widehat{d}_{\text{FQA}}$-based clustering solution obtained using specific values of $C$ and $m$ as input, given by a membership matrix $\boldsymbol U=(u_{ic})$, the Xie-Beni index is defined as}

{
\begin{equation*}
    \operatorname{XBI}(C, m)=\frac{\sum_{i=1}^n \sum_{c=1}^C u_{i c}^2 \left\|\widehat{\boldsymbol{\rho}}^{(i)}-\overline{\widehat{\boldsymbol{\rho}}}^{(c)}\right\|^2}{n \min _{c \neq c^{\prime}}\left\|\overline{\widehat{\boldsymbol{\rho}}}^{(c)}-\overline{\widehat{\boldsymbol{\rho}}}^{(c')}\right\|^2}. 
\end{equation*}}

{Minimizing the numerator in the previous expression, which measures the compactness of the fuzzy partition, is the main goal of the fuzzy $C$-means model based on $\widehat{d}_{\text{FQA}}$ when $m=2$. The denominator measures the degree of separation between groups. Specifically, the smaller the numerator, the more compact the clustering partition, whereas the greater the denominator, the larger the degree of separation between groups. Thus, the Xie-Beni index combines the concepts of compactness and separation so that a lower index value indicates a better clustering solution.}

{Note that the definition of the Xie-Beni index is independent of the $\widehat{d}_{\text{FQA}}$-based clustering algorithm used to obtain $\boldsymbol U$. If the fuzzy $C$-means procedure is employed, the resulting centroids ($\overline{\widehat{\boldsymbol{\rho}}}^{(c)}$, $c=1,\ldots, C$) can be obtained directly from the output of the algorithm. On the contrary, if the fuzzy $C$-medoids method is used, the centroids can still be easily calculated using the corresponding fuzzy solution as input to~\eqref{updatecentroids}, or an alternative definition of the Xie-Beni index using the medoids as prototypes could be considered. Some works on clustering of time series have employed similar criteria to select the parameters $C$ and $m$ \citep{lopez2022quantile1, lopez2023hard}.}  

\clearpage
\section{Two-dimensional scaling representations for the generating processes in Scenarios~1 and~2}\label{section2DS}

As an exploratory step, we carried out a metric two-dimensional scaling (2DS) based on the distance $\widehat{d}_{\text{FQA}}$ in the context of Section 4.2.1 in the main text. This tool is often used to obtain an idea of the capability of a given metric to discriminate between specific underlying groups. Given a pairwise dissimilarity matrix, $\boldsymbol D=(D_{ij})_{1 \leq i,j \leq n}$, a 2DS looks for the set of points $\{(a_i, b_i), i = 1,\ldots, n\}$ minimizing the following expression (called stress):
\begin{equation*}
\sqrt{\frac{\sum_{i \ne j=1}^{n}(\norm{(a_i, b_i)-(a_j, b_j)}-D_{ij})^2}{\sum_{i \ne j = 1}^{n}D_{ij}^2}}.
\end{equation*}

The goal is to represent the distances $D_{ij}$ in terms of Euclidean distances into a two-dimensional space so that the original distances are preserved as well as possible. The lower the value of the stress function, the more reliable the 2DS configuration. This way, a 2DS plot provides a valuable visual representation of how the elements are located with respect to each other according to the original distance.

\begin{figure}[!htb]
\centering
\begin{subfigure}{.5\textwidth}
\centering
\includegraphics[width=0.98\linewidth]{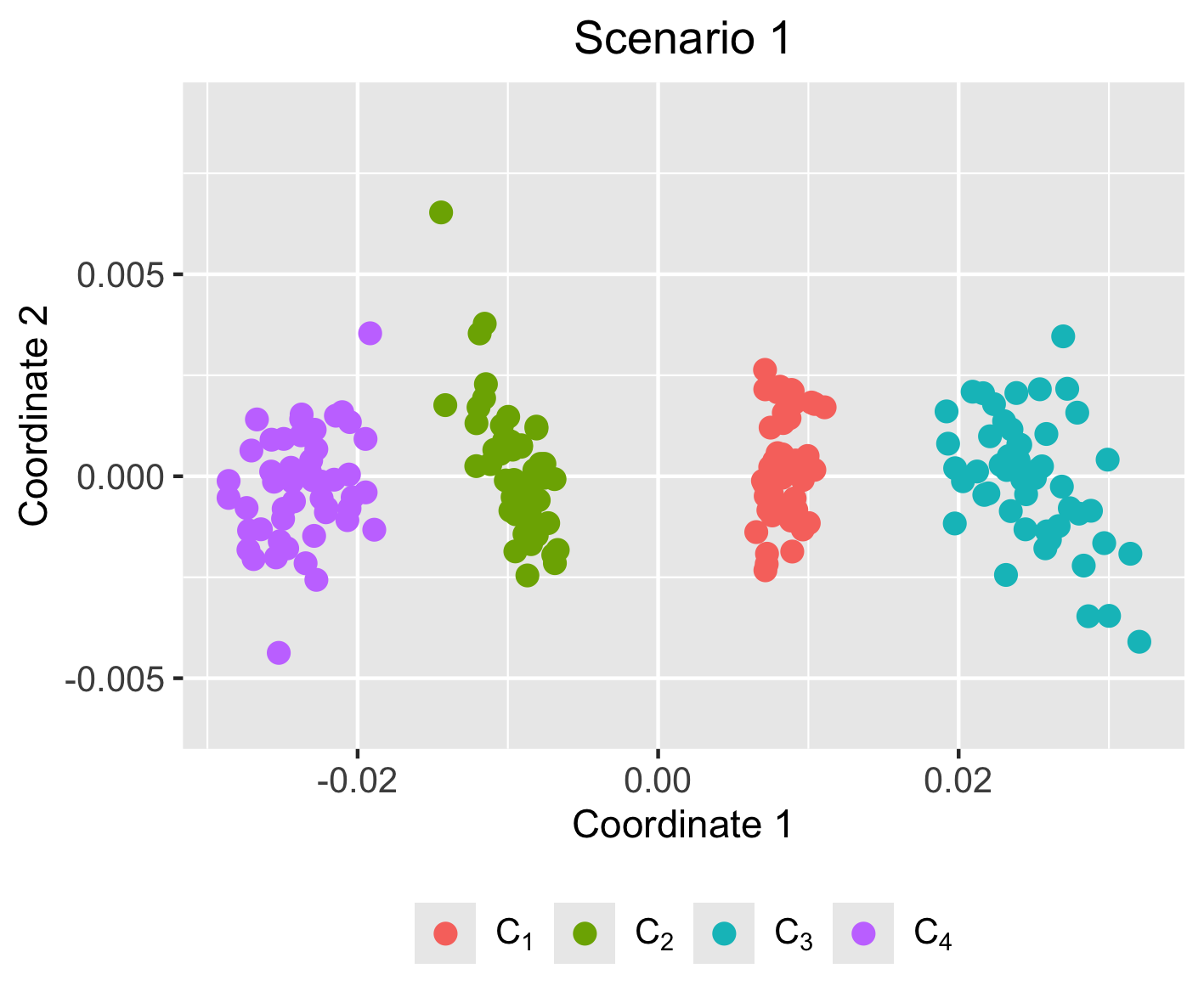}
\end{subfigure}%
\begin{subfigure}{.5\textwidth}
\centering
\includegraphics[width=0.98\linewidth]{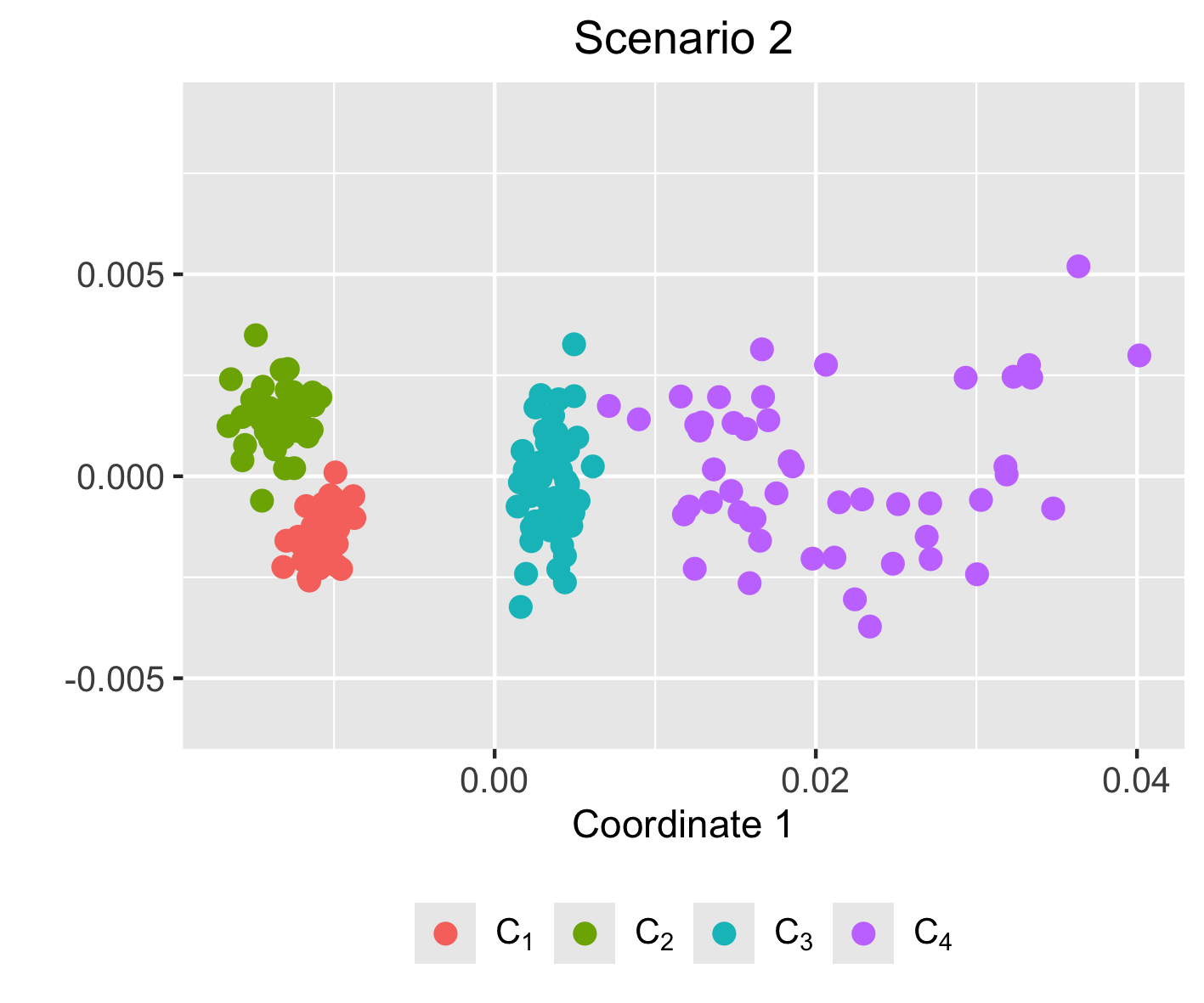}
\end{subfigure}
\caption{\small 2DS planes based on distance $\widehat{d}_{\text{FQA}}$ for the FAR and FGARCH models described in Scenarios~1 and~2 considering series of length $T=1000$.}
\label{Fig_1}
\end{figure}

To obtain informative plots, 50 functional time series of length $T =1000$ were simulated from each generating model. The 2DS was performed for each set of 200 functional time series by calculating the pairwise distance matrix based on $\widehat{d}_{\text{FQA}}$. We considered the set of lags $\mathcal{L}=\{1, 2\}$ in Scenario 1 and $\mathcal{L}=\{1\}$ in Scenario 2. Figure~\ref{Fig_1} contains the corresponding plots, in which the different colors correspond to different generating processes. {To assess the quality of the embeddings, we computed the corresponding stress measures and the $p$-values associated with a permutation test applied to the corresponding dissimilarity matrices \citep{mair2016goodness}. The null hypothesis of this test is that the configuration is obtained from a random permutation of the dissimilarities. We obtained stress measures of 1.26\% and 1.55\% for the 2DS associated with Scenarios 1 and 2, respectively, and $p$-values below 0.01 in both cases. Note that both stress measures are below 5\%, thus indicating an excellent goodness-of-fit \citep{dugard2010approaching}, and both $p$-values are clearly significant. Therefore, the previous results indicate that the graphs in Figure~\ref{Fig_1} provide a reliable summary of the latent representations according to the distance $\widehat{d}_{\text{FQA}}$.}

Different structures can be observed in the 2DS plots of Figure~\ref{Fig_1}. In Scenario~1, the metric can differentiate between the four groups, which was expected due to the different model coefficients in the corresponding FAR processes. In contrast, the 2DS plot for Scenario~2 presents a configuration with a cluster ($\mathcal{C}_4$) the elements of which are substantially spread out, and three more compact groups, thus suggesting a challenging clustering setting. 

\clearpage
\section{{Results for the fuzzy \texorpdfstring{$C$}{C}-means models in the scenarios with well-defined groups}}

The results regarding the fuzzy $C$-means approaches in Scenarios 1 and 2 are provided in Table~\ref{table2as} using a format analogous to one employed in Table~1 in the main text. For the dissimilarity $\widehat{d}_{\text{FQA}}$, the fuzzy $C$-means model results in a slightly lower clustering accuracy than the fuzzy $C$-medoids model for low values of $m$. On the contrary, when $m \in \{1.6, 1.8, 2\}$, the former model generally leads to higher scores. For the remaining metrics, fuzzy $C$-means results in a worse clustering effectiveness in almost all cases. The only exception involves the metric $\widehat{d}_{\text{FACF}}$ in Scenario 2 with $T=200$. Note that, in this scenario, the fuzzy $C$-means model based on $\widehat{d}_{\text{FQA}}$ achieves a significantly higher clustering accuracy than the remaining fuzzy $C$-means approaches in all cases.

\vspace{-.2in}
\begin{center}
\tabcolsep 0.12in
\renewcommand*{\arraystretch}{0.76}
{\begin{longtable}{@{}llllllll@{}}
\caption{{Average ARIF and JIF (in brackets) for different fuzzy $C$-means methods in the crisp scenarios. The highest scores are highlighted in bold. An asterisk indicates that the corresponding method is more effective than the ones without an asterisk for a significance level of 0.01.}}\label{table2as} \\
\toprule 
Model & $T$ &  $m$ & $\widehat{d}_{\text{FQA}}$  & $\widehat{d}_{\text{FACF}}$ & $\widehat{d}_{\text{FSACF}}$  & $\widehat{d}_{\text{K}_\text{m}}$ & $\widehat{d}_{\text{K}_\text{i}}$  \\\midrule
\endfirsthead
\toprule
\endhead
\hline
\multicolumn{8}{r}{{Continued on next page}} \\ 
\endfoot
\endlastfoot
Linear & 200 & $1.2$   & 0.87 (0.83) & 0.41 (0.39) & 0.35 (0.34) & 0.85 (0.81) & \textBF{0.99}$^*$ (\textBF{0.86}$^*$) \\ 
 & & $1.4$ & 0.84 (0.78) & 0.39 (0.37) & 0.32 (0.32) & 0.83 (0.78) & \textBF{0.87}$^*$ (\textBF{0.83}$^*$) \\ 
&   & $1.6$   & 0.78 (0.71) & 0.36 (0.35) & 0.29 (0.30) & 0.81 (0.75) & \textBF{0.85}$^*$ (\textBF{0.80}$^*$) \\ 
&  & $1.8$   & 0.70 (0.63) & 0.34 (0.34) & 0.26 (0.29) & 0.76 (0.70) & \textBF{0.81}$^*$ (\textBF{0.75}$^*$) \\
&  & $2.0$   & 0.62 (0.56) & 0.31 (0.32) & 0.23 (0.28) & 0.72 (0.65) & \textBF{0.76}$^*$ (\textBF{0.69}$^*$) \\
\cmidrule{2-8}
 & 600 & $1.2$   & 0.99$^*$ (\textbf{0.99}$^*$) & 0.54 (0.49) & 0.49 (0.44) & 0.99$^*$ (0.98$^*$) & \textbf{1.00}$^*$ (\textbf{0.99}$^*$) \\ 
&  & $1.4$  & 0.98$^*$ (0.98$^*$) & 0.52 (0.47) & 0.46 (0.42) & 0.98$^*$ (0.97$^*$) & \textbf{0.99}$^*$ (\textbf{0.99}$^*$) \\ 
&  & $1.6$   & 0.96$^*$ (0.94$^*$) & 0.50 (0.46) & 0.43 (0.40) & 0.97$^*$ (0.95$^*$) & \textbf{0.98}$^*$ (\textbf{0.97}$^*$) \\ 
&  & $1.8$ & 0.91 (0.87) & 0.48 (0.44) & 0.40 (0.39) & 0.94$^*$ (0.92$^*$) & \textbf{0.96}$^*$ (\textbf{0.94}$^*$) \\ 
&  & $2.0$  & 0.85 (0.79) & 0.45 (0.42) & 0.37 (0.37) & 0.91$^*$ (0.86$^*$) & \textbf{0.93}$^*$ (\textbf{0.89}$^*$) \\ \midrule
Nonlinear & 200 & $1.2$   & \textBF{0.81}$^*$ (\textBF{0.75}$^*$) & 0.64 (0.59) & 0.40 (0.38) & 0.46 (0.42) & 0.51 (0.46) \\
& & $1.4$ & \textBF{0.78}$^*$ (\textBF{0.72}$^*$) & 0.59 (0.54) & 0.38 (0.36) & 0.44 (0.40) & 0.50 (0.45) \\ 
& & $1.6$   & \textBF{0.73}$^*$ (\textBF{0.66}$^*$) & 0.56 (0.51) & 0.36 (0.35) & 0.42 (0.39) & 0.48 (0.44) \\
& & $1.8$   & \textBF{0.66}$^*$ (\textBF{0.59}$^*$) & 0.53 (0.49) & 0.34 (0.34) & 0.39 (0.38) & 0.45 (0.42) \\ 
& & $2.0$   & \textBF{0.59}$^*$ (\textBF{0.53}$^*$) & 0.50 (0.47) & 0.32 (0.33) & 0.37 (0.36) & 0.43 (0.40) \\ 
\cmidrule{2-8}
& 600 & $1.2$  & \textBF{0.95}$^*$ (\textBF{0.93}$^*$) & 0.82 (0.76) & 0.60 (0.54) & 0.58 (0.52) & 0.66 (0.58) \\ 
& & $1.4$  & \textBF{0.94}$^*$ (\textBF{0.91}$^*$) & 0.79 (0.74) & 0.58 (0.52) & 0.56 (0.50) & 0.64 (0.57) \\ 
& & $1.6$   & \textBF{0.90}$^*$ (\textBF{0.87}$^*$) & 0.77 (0.71) & 0.56 (0.50) & 0.54 (0.48) & 0.63 (0.56) \\ 
& & $1.8$ & \textBF{0.85}$^*$ (\textBF{0.80}$^*$) & 0.74 (0.68) & 0.54 (0.49) & 0.51 (0.47) & 0.61 (0.54) \\ 
& & $2.0$  & \textBF{0.79}$^*$ (\textBF{0.73}$^*$) & 0.70 (0.64) & 0.51 (0.47) & 0.49 (0.45) & 0.58 (0.52) \\  				
\bottomrule
\end{longtable}}
\end{center}

\clearpage
\section{Additional results in the scenarios with well-defined groups}

Note that the practical implementation of $\widehat{d}_{\text{FQA}}$ requires fixing the collection of quantile levels~$\mathcal{T}$ (that also determines the collection of thresholds according to Section~3.2 in the main text), which was set to $\mathcal{T}=\{0.1, 0.5, 0.9\}$ in the original experiments. In this regard, in order to analyze the influence of $\mathcal{T}$ in the performance of the $\widehat{d}_{\text{FQA}}$-based clustering algorithms, we decided to replicate the original simulations considering $\mathcal{T}=\{0.1, 0.3, 0.5, 0.7, 0.9\}$ for the computation of the proposed metric. The average values of the clustering indexes for this extended set of quantile levels are given in Table \ref{table5q} for the fuzzy $C$-medoids model, and in Table \ref{table5q2} for the fuzzy $C$-means model. The results indicate that, in most cases, including more quantile levels does not necessarily lead to better clustering accuracy. This is consistent with several works on quantile-based time series clustering \citep{lafuente2016clustering,vilar2018quantile, lopez2021quantile, lopez2022quantile1}, where it is shown that considering the set $\mathcal{T}=\{0.1, 0.5, 0.9\}$ is usually enough to reach an optimal clustering accuracy.

\begin{table}[!htb]
\centering
\tabcolsep 0.5in
\caption{\small {Average ARIF and JIF (in brackets) for the fuzzy $C$-medoids model based on $\widehat{d}_{\text{FQA}}$ using the set of quantile levels $\mathcal{T}=\{0.1, 0.3, 0.5, 0.7, 0.9\}$.}}\label{table5q}
{\begin{tabular}{@{}lllll@{}}
\toprule 
  & \multicolumn{2}{c}{Linear model} & \multicolumn{2}{c}{Nonlinear model} \\
$m$ & $T=200$  & $T=600$ & $T=200$  & $T=600$   \\\midrule
  $1.2$   & 0.90 (0.86) & 0.99 (0.99) & 0.89 (0.85) & 0.99 (0.99)     \\
  $1.4$ & 0.81 (0.75) & 0.98 (0.97) & 0.81 (0.75) & 0.96 (0.94)   \\
   $1.6$   & 0.71 (0.64) & 0.93 (0.90) & 0.72 (0.65) & 0.90 (0.86)      \\
   $1.8$   & 0.61 (0.55) & 0.87 (0.82) & 0.62 (0.56) & 0.83 (0.76)      \\
   $2.0$   & 0.53 (0.48) & 0.79 (0.73) & 0.54 (0.49) & 0.75 (0.68)     \\
\bottomrule
\end{tabular}}
\end{table}

\begin{table}[!htb]
\centering
\tabcolsep 0.48in
\caption{{Average ARIF and JIF (in brackets) for the fuzzy $C$-means model based on $\widehat{d}_{\text{FQA}}$ using the set of quantile levels $\mathcal{T}=\{0.1, 0.3, 0.5, 0.7, 0.9\}$.}}\label{table5q2}
{\begin{tabular}{@{}lllll@{}}
\toprule 
  & \multicolumn{2}{c}{Linear model} & \multicolumn{2}{c}{Nonlinear model} \\
$m$ & $T=200$  & $T=600$ & $T=200$  & $T=600$   \\\midrule
  $1.2$   & 0.86 (0.82) & 0.99 (0.99) & 0.81 (0.76) & 0.95 (0.93)    \\
   $1.4$  & 0.83 (0.78) & 0.98 (0.98) & 0.79 (0.73) & 0.93 (0.91)    \\
   $1.6$   & 0.78 (0.71) & 0.96 (0.94) & 0.74 (0.67) & 0.90 (0.86)       \\
   $1.8$ & 0.70 (0.64) & 0.91 (0.87) & 0.72 (0.64) & 0.85 (0.80)     \\
   $2.0$  & 0.63 (0.56) & 0.85 (0.80) & 0.68 (0.61) & 0.79 (0.72)      \\
\bottomrule
\end{tabular}}
\end{table}

Since one of the advantages of the clustering procedures based on $\widehat{d}_{\text{FQA}}$ is that they can deal with functional time series with unequal lengths, we decided to perform additional simulations involving sets of time series with different values for $T$. We considered again four groups of five functional time series each but, this time, the length of each time series was determined by randomly selecting a value from the set $\{200, 300, 400, 500, 600\}$, with all the values having the same probability ($1/5$). The remaining simulation elements were kept the same as in the original experiments. The corresponding results are given in Table~\ref{tableaa} for the fuzzy $C$-medoids models, and in Table \ref{tableaa2} for the fuzzy $C$-means models, and are consistent with the original results. In Scenario~1, the metrics $\widehat{d}_{\text{K}_\text{m}}$ and $\widehat{d}_{\text{K}_\text{i}}$ are associated with the best scores, but the proposed distance also shows a high clustering accuracy. In Scenario~2, both the fuzzy $C$-medoids and the fuzzy $C$-means procedures based on $\widehat{d}_{\text{FQA}}$ achieve a substantially higher clustering accuracy than their counterparts for all values of $m$.

\begin{table}[!htb]
\centering
\tabcolsep 0.16in
\caption{\small {Average ARIF and JIF (in brackets) for different fuzzy $C$-medoids methods in the crisp scenarios considering sets of time series with different lengths. The highest scores are highlighted in bold. An asterisk indicates that the corresponding method is more effective than the ones without an asterisk for a significance level of 0.01.}}\label{tableaa}
{\begin{tabular}{@{}lllllll@{}}
\toprule 
 Model &  $m$ & $\widehat{d}_{\text{FQA}}$  & $\widehat{d}_{\text{FACF}}$ & $\widehat{d}_{\text{FSACF}}$  & $\widehat{d}_{\text{K}_\text{m}}$ & $\widehat{d}_{\text{K}_\text{i}}$  \\\midrule
 Linear & $1.2$   & 0.97 (0.96) & 0.56 (0.51) & 0.48 (0.44) & \textbf{0.99}$^*$ (\textbf{0.99}$^*$) & \textbf{0.99}$^*$ (\textbf{0.99}$^*$) \\ 
 & $1.4$ & 0.92 (0.89) & 0.52 (0.48) & 0.44 (0.42) & 0.96$^*$ (0.94$^*$) & \textbf{0.97}$^*$ (\textbf{0.96}$^*$) \\ 
 & $1.6$   & 0.85 (0.79) & 0.48 (0.45) & 0.40 (0.39) & 0.93 (0.90) & \textbf{0.95}$^*$ (\textbf{0.92}$^*$) \\ 
& $1.8$   & 0.76 (0.69) & 0.45 (0.42) & 0.36 (0.37) & 0.88 (0.83) & \textbf{0.90}$^*$ (\textbf{0.86}$^*$) \\ 
 & $2.0$   & 0.68 (0.61) & 0.41 (0.40) & 0.33 (0.35) & 0.83 (0.77) & \textbf{0.85}$^*$ (\textbf{0.80}$^*$) \\ 
\midrule
 Nonlinear & $1.2$   & \textbf{0.94}$^*$ (\textbf{0.91}$^*$) & 0.71 (0.65) & 0.60 (0.54) & 0.62 (0.55) & 0.67 (0.59) \\ 
 & $1.4$ & \textbf{0.88}$^*$ (\textbf{0.83}$^*$) & 0.68 (0.61) & 0.57 (0.51) & 0.58 (0.52) & 0.64 (0.57) \\ 
 & $1.6$   & \textbf{0.80}$^*$ (\textbf{0.74}$^*$) & 0.65 (0.58) & 0.53 (0.48) & 0.54 (0.49) & 0.61 (0.55) \\ 
 & $1.8$   & \textbf{0.72}$^*$ (\textbf{0.66}$^*$) & 0.62 (0.55) & 0.50 (0.46) & 0.50 (0.46) & 0.58 (0.52) \\  
 & $2.0$   & \textbf{0.65}$^*$ (\textbf{0.58}$^*$) & 0.58 (0.53) & 0.47 (0.44) & 0.46 (0.43) & 0.54 (0.49) \\ 		
\bottomrule
\end{tabular}}
\end{table}

\begin{table}[!htb]
\centering
\tabcolsep 0.17in
\caption{{Average ARIF and JIF (in brackets) for different fuzzy $C$-means methods in the crisp scenarios considering sets of time series with different lengths. The highest scores are highlighted in bold. An asterisk indicates that the corresponding method is more effective than the ones without an asterisk for a significance level of 0.01.}}\label{tableaa2}
{\begin{tabular}{@{}lllllll@{}}
\toprule 
 Model &  $m$ & $\widehat{d}_{\text{FQA}}$  & $\widehat{d}_{\text{FACF}}$ & $\widehat{d}_{\text{FSACF}}$  & $\widehat{d}_{\text{K}_\text{m}}$ & $\widehat{d}_{\text{K}_\text{i}}$  \\\midrule
 Linear & $1.2$   & 0.95$^*$ (0.93$^*$) & 0.49 (0.44) & 0.43 (0.40) & 0.93 (0.91) & \textbf{0.96}$^*$ (\textbf{0.94}$^*$) \\
  & $1.4$  & 0.93 (0.90) & 0.46 (0.43) & 0.40 (0.38) & 0.92 (0.90) & \textbf{0.95}$^*$ (\textbf{0.93}$^*$) \\
  & $1.6$   & 0.89 (0.84) & 0.44 (0.41) & 0.37 (0.36) & 0.91 (0.87) & \textbf{0.93}$^*$ (\textbf{0.90}$^*$) \\ 
  & $1.8$ & 0.82 (0.76) & 0.41 (0.39) & 0.34 (0.34) & 0.87 (0.83) & \textbf{0.90}$^*$ (\textbf{0.86}$^*$) \\
  & $2.0$  & 0.75 (0.68) & 0.38 (0.37) & 0.31 (0.32) & 0.83 (0.77) & \textbf{0.86}$^*$ (\textbf{0.80}$^*$) \\ \midrule
 Nonlinear & $1.2$  & \textbf{0.89}$^*$ (\textbf{0.85}$^*$) & 0.64 (0.57) & 0.51 (0.46) & 0.53 (0.48) & 0.62 (0.55) \\ 
 & $1.4$  & \textbf{0.87}$^*$ (\textbf{0.82}$^*$)  & 0.63 (0.56) & 0.49 (0.45) & 0.51 (0.46) & 0.59 (0.53) \\ 
& $1.6$   & \textbf{0.83}$^*$ (\textbf{0.77}$^*$) & 0.61 (0.54) & 0.47 (0.43) & 0.49 (0.44) & 0.57 (0.49) \\ 
& $1.8$ & \textbf{0.77}$^*$ (\textbf{0.70}$^*$) & 0.58 (0.52) & 0.45 (0.42) & 0.46 (0.43) & 0.55 (0.49) \\ 
& $2.0$  & \textbf{0.70}$^*$ (\textbf{0.64}$^*$) & 0.56 (0.50) & 0.43 (0.40) & 0.44 (0.41) & 0.52 (0.47) \\ 		
\bottomrule
\end{tabular}}
\end{table}

\clearpage
\section{{Results for the fuzzy \texorpdfstring{$C$}{C}-means models in the scenarios with some degree of uncertainty}}

\begin{figure}[!htb]
\centering
\includegraphics[width=\textwidth]{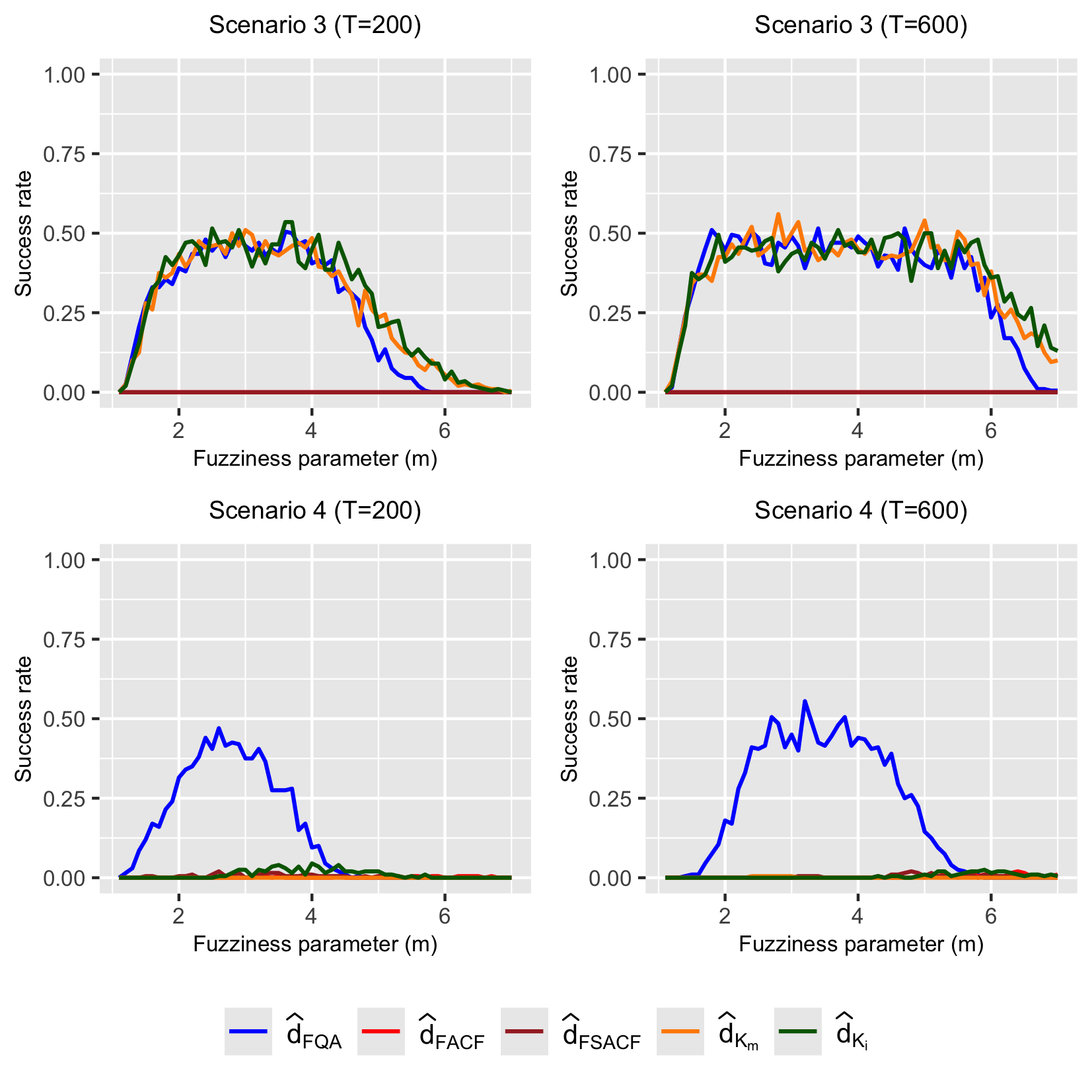}
\caption{Success rates with respect to $m$ for the fuzzy $C$-means model based on different metrics in the uncertain scenarios.}
\label{figfcmdu}
\end{figure}

Figure \ref{figfcmdu} shows the rates of correct classification as a function of $m$ for the fuzzy $C$-means procedure based on several metrics in Scenarios 3 and 4. The corresponding maximum value and area under the fuzziness curve are provided in Table \ref{tablefcmdu}.

\begin{table}[!htb]
\centering
\tabcolsep 0.45in
\caption{\small {Maximum value and area under the fuzziness curve (in brackets) for the fuzzy $C$-means model based on several dissimilarities in the uncertain scenarios. The highest scores are highlighted in bold.}}\label{tablefcmdu}
{\begin{tabular}{@{}lllll@{}}
\toprule 
               & \multicolumn{2}{c}{Scenario 3} & \multicolumn{2}{c}{Scenario 4} \\
 Metric  & $T=200$  & $T=600$ & $T=200$  & $T=600$   \\\midrule
 $\widehat{d}_{\text{FQA}}$   & 0.51 (1.48)  & 0.52 (2.13) & \textbf{0.47} (\textbf{0.82}) &  \textbf{0.56} (\textbf{1.24})   \\
  $\widehat{d}_{\text{FACF}}$  & 0.00 (0.00) & 0.00 (0.00) & 0.01 (0.01) & 0.02 (0.02)    \\
  $\widehat{d}_{\text{FSACF}}$   & 0.00 (0.00) & 0.00 (0.00) &  0.02 (0.02) &  0.02 (0.02)      \\
   $\widehat{d}_{\text{K}_\text{m}}$ & 0.51 (1.61) & \textbf{0.56} (2.25) & 0.00 (0.00) &  0.01 (0.01)    \\
   $\widehat{d}_{\text{K}_\text{i}}$  & \textbf{0.54} (\textbf{1.69}) & 0.51 (\textbf{2.29}) & 0.05 (0.06) &   0.03 (0.03)    \\
\bottomrule
\end{tabular}}
\end{table}

\clearpage

\section{Computational time assessment} \label{subsectiontimeconsupmtion}

In order to analyze the efficiency of the clustering approaches examined throughout Section~4 in the main text, we present the runtime of the corresponding programs used for the experiments in Scenario~1. {For all the methods except for $\widehat{d}_{\text{TSY}}$, we consider the runtime associated with the fuzzy $C$-medoids approach}. Given a method and a value for $T$, we report the CPU runtime required to complete the clustering task (for the five values of $m$) for the 200 simulation trials. The computer used to run the programs was a MacBook Pro with an Apple M2 Pro chip and 16 GB of RAM. The programs were coded and executed in the \Rlogo\ software \citep{rjournal} (version 4.3.1). 

The runtime (minutes) associated with each method is provided in Table~\ref{tablecomputationtimes}. The fuzzy $C$-medoids method based on $\widehat{d}_{\text{FQA}}$ is associated with the lowest runtime for both values of $T$. In contrast, $\widehat{d}_{\text{TSY}}$ is by far associated with the highest runtime. Additionally, the computation time of the proposed approach barely increases with $T$. On the contrary, the runtime associated with distances $\widehat{d}_{\text{K}_\text{m}}$ and $\widehat{d}_{\text{K}_\text{i}}$ shows a substantial increase with $T$, which is expected, since the feature extraction mechanism of these metrics entails a number of $\sum_{k=1}^{L}\binom{T-l_k}{2}$ comparisons. Among these two dissimilarities, $\widehat{d}_{\text{K}_\text{i}}$ is computationally heavier due to the integration involved in the definition of $\prec_i$. It may not be feasible to run the method based on $\widehat{d}_{\text{K}_\text{i}}$ in databases containing many time series of moderate to large lengths. {The runtime associated with the dissimilarities $\widehat{d}_{\text{FACF}}$ and $\widehat{d}_{\text{FSACF}}$ increases linearly with $T$, but the former metric implies considerably heavier computations than the latter.}
\begin{table}[!htb]
\centering
\tabcolsep 1.3in
\caption{\small Runtime (minutes) for the different methods regarding the 200 simulation trials in Scenario 1.}\label{tablecomputationtimes}
\begin{tabular}{@{}lrr@{}} 
\toprule
Method   & $T=200$ & $T=600$ \\\midrule
$\widehat{d}_{\text{FQA}}$  & 16.57  & 20.03    \\
{$\widehat{d}_{\text{FACF}}$} & {179.78} & {578.35} \\
{$\widehat{d}_{\text{FSACF}}$}  & {18.42} & {48.34} \\
$\widehat{d}_{\text{K}_\text{m}}$ & 26.17 & 212.81 \\
$\widehat{d}_{\text{K}_\text{i}}$  & 170.34 & 1545.17 \\
$\widehat{d}_{\text{TSY}}$    & 4590.14 & 32906.82 \\\bottomrule
\end{tabular}
\end{table}

The above analyses show that the proposed clustering method is computationally efficient and can be feasibly executed even with databases containing long time series. {For all methods except for $\widehat{d}_{\text{TSY}}$, an analogous time consumption analysis was carried out considering the fuzzy $C$-means approach instead of the fuzzy $C$-medoids. The corresponding results are almost identical to those in Table~\ref{tablecomputationtimes}, since, in all cases, most of the runtime is spent on the feature extraction stage.}

\clearpage 
\section{Two-dimensional scaling representation for the financial application}

\begin{figure}[!htb]
\centering
\includegraphics[width=0.55\textwidth]{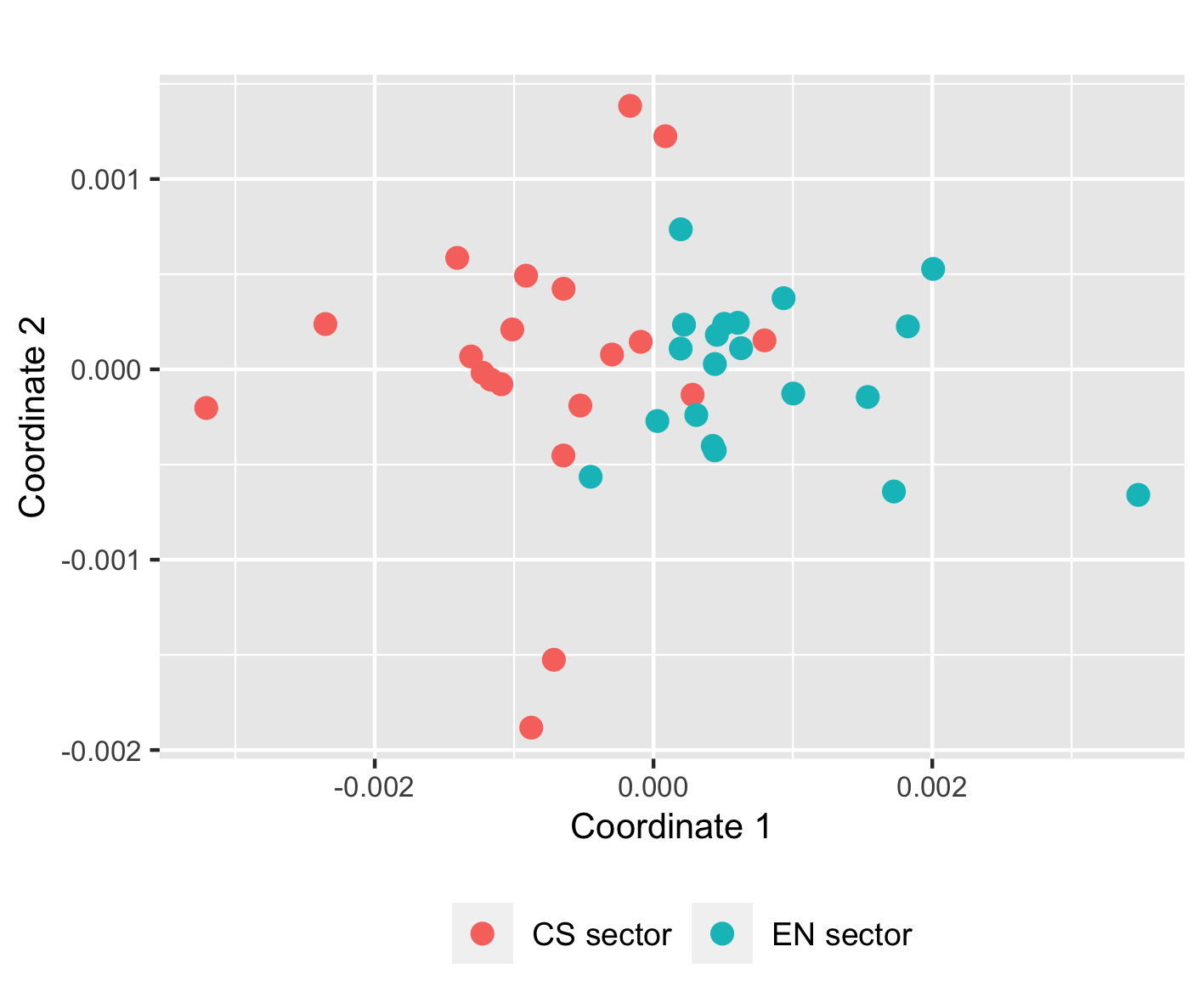}
\caption{\small 2DS plane based on distance $\widehat{d}_{\text{FQA}}$ for the 40 functional time series of log-returns.}
\label{Fig_5}
\end{figure}

In the context of Section 5.1 in the main text, as an exploratory exercise, we constructed a 2DS plot based on $\widehat{d}_{\text{FQA}}$, which was computed considering the chosen values for $\mathcal{T}$ and $\mathcal{L}$. This allowed us to obtain a $\widehat{d}_{\text{FQA}}$-based representation of the 40 financial series in a two-dimensional plane. The corresponding plot is shown in Figure~\ref{Fig_5}, where the points are colored according to the sector associated with each company. {The goodness-of-fit of the 2DS was evaluated through the stress measure and the $p$-value for the permutation test applied to the corresponding dissimilarity matrix (as in Section \ref{section2DS}), which resulted 5.89\% and less than 0.01, respectively. Note that the stress measure is very close to 5\%, the threshold below which the 2DS representation is considered satisfactory \citep{dugard2010approaching}, and the $p$-value is clearly significant. Therefore, the plot in Figure~\ref{Fig_5} provides a reliable representation of the underlying structure according to the distance $\widehat{d}_{\text{FQA}}$.}

The 2DS plot indicates that the distance $\widehat{d}_{\text{FQA}}$ can identify a certain structure related to the considered sectors, since the points associated with CS and EN companies are concentrated on the left and right parts of the graph, respectively. However, there is a moderate degree of overlap between the two groups. In addition,
we can see a few series in each cluster displaying a dependence pattern that differs from that shown by most of the time series in their associated group. Previous comments suggest that the fuzzy paradigm could be suitable for clustering in the considered database. 

\clearpage 
\section{Clustering age-specific mortality improvement rates}

We consider age-specific mortality rates in 41 countries over several years. The mortality rates are the ratios of death counts to population exposure in the relevant year for a given age interval. We examine age groupings ranging from 0 to 110 in single years of age, with the last age group including all ages of 110 and above. The data were sourced from the \cite{HMD24}. As the available sample period differs among the 41 countries, and there are some countries with only a few years of available data, we decided to consider only the countries for which we have mortality data since at least 1960. For such countries, our sample period is from 1960 till the last year of available data, which varies among countries. Among the original countries, 33 were selected, namely Australia (AUS), Austria (AUT), Belarus (BLR), Belgium (BEL), Bulgaria (BGR), Canada (CAN), Czech Republic (CZE), Denmark (DNK), Estonia (EST), Finland (FIN), France (FRA), Hungary (HUN), Iceland (ISL), Ireland (IRL), Italy (ITA), Japan (JPN), Latvia (LVA), Lithuania (LTU), Luxembourg (LUX), the Netherlands (NLD), New Zealand (NZL), Norway (NOR), Poland (POL), Portugal (PRT), Russia (RUS), Slovakia (SVK), Spain (ESP), Sweden (SWE), Switzerland (CHE), Taiwan (TWN), the United Kingdom (GBR), Ukraine (UKR), and the United States (USA).

\begin{figure}[!htb]
\centering
\begin{subfigure}{.5\textwidth}
\centering
\includegraphics[width=0.98\linewidth]{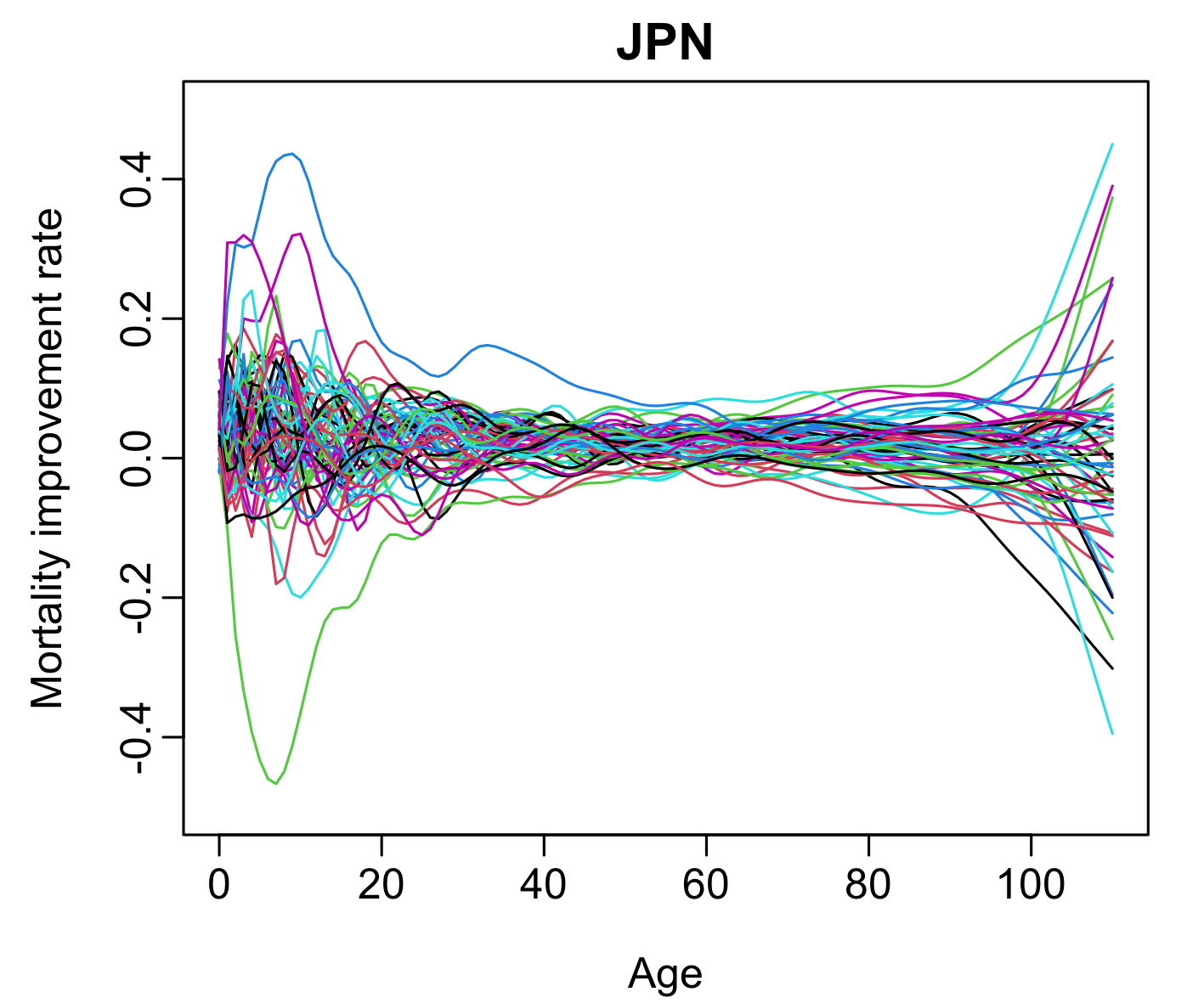}
\end{subfigure}%
\begin{subfigure}{.5\textwidth}
\centering
\includegraphics[width=0.98\linewidth]{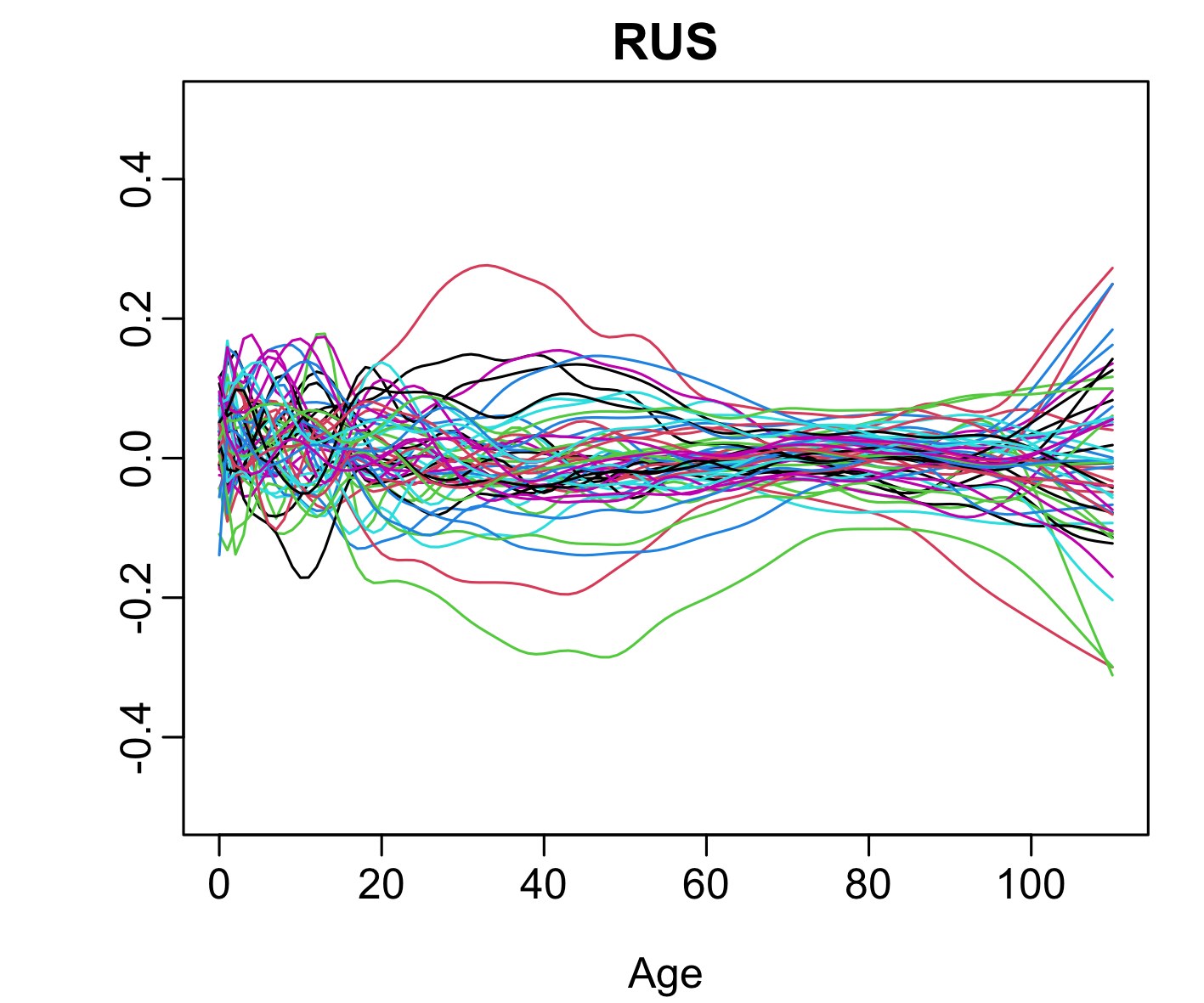}
\end{subfigure}
\caption{\small Functional time series of mortality improvement rates for JPN (from 1960 to 2022) and RUS (from 1960 to 2014).}\label{Fig_6}
\end{figure}

The mortality data from the selected countries were used to construct a collection of 33 functional time series by considering the yearly curves of mortality rates as a function of age. As the 33 countries are associated with different ending years, our series have different lengths, which vary from $T=54$ (UKR) to $T=64$ (DNK). These series were smoothed by using the function \textit{smooth.demogdata()} of the \Rlogo\ package \textbf{demography} \citep{demography}, tailored to handle any missing value or poor data quality issues at the higher ages. As the smoothed functional time series of mortality rates are not stationary, the series was transformed into so-called mortality improvement rates \citep{HR12, Shang19}. Specifically, if $M_{i,t}(u_j)$ is the smoothed mortality rate of the $i$\textsuperscript{th} country for age $u_j$ on year $t$, the series of mortality improvement rates is defined as 
\begin{equation*}
M^{*}_{i,t}\left(u_j\right)=2\times\frac{M_{i,t-1}(u_j)-M_{i,t}(u_j)}{M_{i,t-1}(u_j)+M_{i,t}(u_j)},
\end{equation*}
where $i=1,2, \ldots, 33$, $j=0,1, \ldots, 110$, and $t=2,3, \ldots, L_i$, being $L_i$ the length of the series of mortality rates for the $i$\textsuperscript{th} country. For demonstration, the functional time series of mortality improvement rates for JPN and RUS are displayed in Figure~\ref{Fig_6}. 

We applied the fuzzy $C$-medoids and the fuzzy $C$-means models based on $\widehat{d}_{\text{FQA}}$ to the dataset of 33 functional time series of mortality improvement rates. As in the first application, and, according to Section~3.2 in the main text, the set $\mathcal{T}$ was selected as $\mathcal{T}=\{0.1, 0.5, 0.9\}$. The collection of lags, $\mathcal{L}$, was set to $\mathcal{L}=\{1\}$. This time, as the number of clusters was not fixed in advance, the values of $C$ and $m$ were selected simultaneously through the criterion based on internal clustering quality indexes described in Section~3.2.3 in the main text, resulting in $C=2$ and $m=1.5$ {(the selection was again performed for the fuzzy $C$-means model and the resulting values of $C$ and $m$ were also used in the fuzzy $C$-medoids model for the sake of comparison)}. The grids considered for the selection of $\mathcal{L}$ and $m$ were the same as in the financial application. The grid associated with the number of groups was given by the set $\{2, \ldots, 6\}$.  


The clustering partition produced by the fuzzy $C$-medoids algorithm is given in the left part of Table~\ref{tablefuzzycountries}, where the bold font has the same meaning as in the above application. As expected from the selected value for $m$, the highest membership degree is close to one for almost all countries, thus suggesting an underlying clustering configuration with a low degree of fuzziness. Cluster $\mathcal{C}_1$ contains many European countries along with AUS, NZL, and TWN. Cluster $\mathcal{C}_2$ includes BLR, RUS, and UKR, along with the North American countries CAN and USA, the Baltic states EST, LVA, and LTU, and SVK. It is worth highlighting that this second group makes sense from a geographical and political point of view. For instance, until 1991, the six neighboring countries BLR, RUS, UKR, EST, LVA, and LTU belonged to the same state, the Soviet Union. Therefore, it is not surprising that the mortality improvement rates in these countries show similar behavior over the last 60 years. The only country with membership degrees substantially spread out is JPN, the mortality improvement rates of which share some patterns with the series in both groups. {The clustering partition associated with the fuzzy $C$-means procedure is provided in the left part of Table~\ref{tablefuzzycountries2}.} 
\begin{center}
\tabcolsep 0.175in
\begin{longtable}{@{}lll|ll|llllll@{}} 
\caption{\small Fuzzy solutions produced by the fuzzy $C$-medoids procedure based on $\widehat{d}_{\text{FQA}}$ when grouping the 33 functional time series of mortality improvement rates. The left, middle, and right parts show the partitions associated with the total, male, and female populations, respectively. Membership degrees above 0.7 are written in bold.} \label{tablefuzzycountries} \\\toprule
& \multicolumn{2}{c}{Total} & \multicolumn{2}{c}{Male} & \multicolumn{6}{c}{Female}   \\ \cmidrule{2-11}
Country    & $\mathcal{C}_1$  & $\mathcal{C}_2$ & $\mathcal{C}_1$  & $\mathcal{C}_2$ & $\mathcal{C}_1$  & $\mathcal{C}_2$ & $\mathcal{C}_3$  & $\mathcal{C}_4$ & $\mathcal{C}_5$  & $\mathcal{C}_6$  \\\midrule
\endfirsthead
\toprule
& \multicolumn{2}{c}{Total} & \multicolumn{2}{c}{Male} & \multicolumn{6}{c}{Female}  \\\cmidrule{2-11}
Country    & $\mathcal{C}_1$  & $\mathcal{C}_2$ & $\mathcal{C}_1$  & $\mathcal{C}_2$ & $\mathcal{C}_1$  & $\mathcal{C}_2$  & $\mathcal{C}_3$  & $\mathcal{C}_4$  & $\mathcal{C}_5$  & $\mathcal{C}_6$  \\\midrule
\endhead
\hline \multicolumn{11}{r}{{Continued on next page}} \\
\endfoot
\endlastfoot
AUS &       \textBF{0.88} & 0.12 & \textBF{0.98} & 0.02 & 0.07 & 0.05 & 0.10 & 0.05 & 0.21 & 0.52 \\
AUT &  \textBF{0.88} & 0.12 & \textBF{0.94} & 0.06 & 0.36 & 0.12 & 0.12 & 0.02 & 0.34 & 0.04   \\      
BLR  &     0.22 & \textBF{0.78} & \textBF{0.97} & 0.03 & 0.01 & \textBF{0.91} & 0.01 & 0.01 & 0.04 & 0.02 \\  
BEL  & \textBF{0.92} & 0.08 & \textBF{0.94} & 0.06 & 0.08 & 0.04 & 0.17 & 0.04 & 0.33 & 0.34 \\       
BGR &  \textBF{0.93} &  0.07 & \textBF{0.98} & 0.02 & 0.00 & 0.00 & 0.00 & 0.00 & \textBF{1.00} & 0.00 \\        
CAN &  0.10  &  \textBF{0.90} & \textBF{0.86} & 0.14 & 0.05 & 0.58 & 0.07 & 0.05 & 0.09 & 0.16  \\  
CZE &  \textBF{0.86}  &   0.14 & \textBF{1.00} & 0.00 & 0.28 & 0.05 & 0.48 & 0.02 & 0.09 & 0.08 \\     
DNK &  \textBF{0.95} & 0.05 & \textBF{0.97} & 0.03 & 0.31 & 0.07 & 0.05 & 0.01 & 0.29 & 0.27 \\     
EST &  0.25 & \textBF{0.75} & \textBF{0.96} & 0.04 & 0.09 & 0.46 & 0.30 & 0.02 & 0.07 & 0.06 \\        
FIN &  \textBF{0.93} & 0.07 & \textBF{0.99} & 0.01 & 0.11 & 0.03 & 0.18 & 0.04 & 0.18 & 0.46  \\
FRA  &  \textBF{0.96} &  0.04 & \textBF{0.84} & 0.16 & 0.08 & 0.08 & 0.09 & 0.23 & 0.46 & 0.06 \\        
HUN &   \textBF{0.80} & 0.20 & \textBF{0.99} & 0.01 & 0.00 & 0.00 & 0.00 & 0.00 & 0.00 & \textBF{1.00}  \\
ISL  &  0.68  &  0.32  & 0.00 & \textBF{1.00} & 0.12 & 0.14 & 0.14 & 0.11 & 0.24 & 0.25 \\      
IRL & \textBF{0.89}  &  0.11 & \textBF{0.95} & 0.05 & 0.04 & 0.05 & 0.22 & 0.04 & 0.12 & 0.53 \\    
ITA &  \textBF{0.93}  &  0.07 & \textBF{0.96} & 0.04 & 0.37 & 0.05 & 0.20 & 0.02 & 0.29 & 0.07 \\
JPN  & 0.53  & 0.47 & \textBF{0.99} & 0.01 & 0.00 & \textBF{1.00} & 0.00 & 0.00 & 0.00 & 0.00 \\    
LVA    &  0.06 & \textBF{0.94} & \textBF{0.95} & 0.05 & 0.33 & 0.14 & 0.43 & 0.06 & 0.03 & 0.01  \\ 
LTU  & 0.18 & \textBF{0.82} & \textBF{0.92} & 0.08  & 0.14 & 0.33 & 0.30 & 0.17 & 0.03 & 0.03     \\
LUX  &  \textBF{0.93} & 0.07 & \textBF{0.87} & 0.13 & 0.22 & 0.17 & 0.17 & 0.04 & 0.30 & 0.10   \\
NLD      & \textBF{0.92} & 0.08 & \textBF{0.99} & 0.01  & 0.12 & 0.04 & 0.10 & 0.17 & 0.55 & 0.02   \\
NZL  & \textBF{0.93} & 0.07 & \textBF{1.00} & 0.00 & 0.25 & 0.09 & 0.25 & 0.08 & 0.20 & 0.13  \\
NOR  & \textBF{0.78} & 0.22 & \textBF{0.97} & 0.03 & 0.00 & 0.00 & 0.00 & \textBF{1.00} & 0.00 & 0.00  \\
POL     & \textBF{0.96} & 0.04  & \textBF{0.96} & 0.04 & 0.02 & 0.20 & 0.03 & 0.11 & 0.26 & 0.38   \\
PRT    & \textBF{1.00} & 0.00    & \textBF{0.98} & 0.02 & 0.21 & 0.19 & 0.14 & 0.21 & 0.15 & 0.10  \\
RUS   & 0.12 & \textBF{0.88} & \textBF{0.82} & 0.18  & 0.08 & 0.49 & 0.11 & 0.08 & 0.16 & 0.08 \\
SVK   & 0.25 & \textBF{0.75}  & \textBF{0.99} & 0.01 & 0.00 & 0.00 & \textBF{1.00} & 0.00 & 0.00 & 0.00  \\
ESP  & \textBF{0.96} & 0.04 & \textBF{0.98} & 0.02 & 0.06 & 0.25 & 0.20 & 0.26 & 0.21 & 0.02  \\
SWE    & \textBF{0.94} & 0.06 & \textBF{0.99} & 0.01 & \textBF{1.00} & 0.00 & 0.00 & 0.00 & 0.00 & 0.00  \\
CHE  & \textBF{0.89} & 0.11 & \textBF{0.99} & 0.01 & 0.29 & 0.07 & 0.42 & 0.12 & 0.04 & 0.06  \\
TWN & \textBF{0.97} & 0.03   & \textBF{0.95} & 0.05 & 0.46 & 0.09 & 0.04 & 0.09 & 0.29 & 0.03 \\
GBR   & \textBF{0.95} & 0.05 & \textBF{0.99} & 0.01 & 0.05 & 0.03 & 0.07 & 0.08 & 0.52 & 0.25 \\
UKR    & 0.01 & \textBF{0.99} & \textBF{0.87} & 0.13 & 0.02 & \textBF{0.81} & 0.06 & 0.07 & 0.03 & 0.01 \\
USA   & 0.00 & \textBF{1.00}  & \textBF{0.96} & 0.04  & 0.01 & \textBF{0.83} & 0.13 & 0.01 & 0.01 & 0.01  \\
\bottomrule
\end{longtable}
\end{center}

\vspace{-.2in}

\vspace{-.2in}
{\begin{center}
\tabcolsep 0.175in
\begin{longtable}{@{}lll|ll|llllll@{}} 
\caption{{Fuzzy solutions produced by the fuzzy $C$-means procedure based on $\widehat{d}_{\text{FQA}}$ when grouping the 33 functional time series of mortality improvement rates. The left, middle, and right parts show the partitions associated with the total, male and female populations, respectively. Membership degrees above 0.7 are written in bold.}}\label{tablefuzzycountries2} \\\toprule
& \multicolumn{2}{c}{Total} & \multicolumn{2}{c}{Male} & \multicolumn{6}{c}{Female}   \\ \cline{2-11}
Country    & $\mathcal{C}_1$  & $\mathcal{C}_2$ & $\mathcal{C}_1$  & $\mathcal{C}_2$ & $\mathcal{C}_1$  & $\mathcal{C}_2$ & $\mathcal{C}_3$  & $\mathcal{C}_4$ & $\mathcal{C}_5$  & $\mathcal{C}_6$  \\\midrule
\endfirsthead
\toprule
& \multicolumn{2}{c}{Total} & \multicolumn{2}{c}{Male} & \multicolumn{6}{c}{Female}  \\\cline{2-11}
Country    & $\mathcal{C}_1$  & $\mathcal{C}_2$ & $\mathcal{C}_1$  & $\mathcal{C}_2$ & $\mathcal{C}_1$  & $\mathcal{C}_2$  & $\mathcal{C}_3$  & $\mathcal{C}_4$  & $\mathcal{C}_5$  & $\mathcal{C}_6$  \\\midrule
\endhead
\hline \multicolumn{11}{r}{{Continued on next page}} \\
\endfoot
\endlastfoot
AUS &       \textBF{0.91} & 0.09 & \textBF{0.93} & 0.07 & 0.01 & 0.01 & 0.01 & 0.01 & 0.01 & \textbf{0.95} \\
AUT &  \textBF{0.94} & 0.06 & \textBF{0.93} & 0.07 & 0.01 & 0.13 & 0.06 & 0.06 & \textbf{0.73} & 0.01 \\     
BLR  &     0.05 & \textBF{0.95} & 0.14 & \textbf{0.86} & 0.01 & \textbf{0.93} & 0.01 & 0.02 & 0.02 & 0.01 \\
BEL  & \textBF{0.97} & 0.03 & \textBF{0.96} & 0.04 & 0.01 & 0.01 & 0.01 & 0.01 & 0.01 & \textbf{0.95} \\       
BGR &  \textBF{0.94} &  0.06 & \textBF{0.88} & 0.12 & 0.01 & 0.02 & 0.01 & 0.19 & \textbf{0.76} & 0.01 \\        
CAN &  0.02  &  \textBF{0.98} & 0.05 & \textbf{0.95} & 0.01 & \textbf{0.93} & 0.01 & 0.02 & 0.02 & 0.01 \\ 
CZE &  \textBF{0.93}  &   0.07 & \textBF{0.93} & 0.07 & 0.01 & 0.02 & 0.11 & 0.02 & \textbf{0.83} & 0.01 \\   
DNK &  \textBF{0.90} & 0.10 & \textBF{0.95} & 0.05 & 0.01 & 0.06 & 0.03 & 0.02 & \textbf{0.87} & 0.01 \\     
EST &  0.20 & \textBF{0.80} & 0.02 & \textBF{0.98} & 0.01 & \textbf{0.78} & 0.11 & 0.02 & 0.07 & 0.01 \\        
FIN &  \textBF{0.98} & 0.02 & \textBF{0.90} & 0.10 & 0.00 & 0.00 & 0.00 & 0.00 & 0.00 & \textbf{1.00} \\
FRA  &  \textBF{0.99} &  0.01 & \textBF{0.83} & 0.17 & 0.00 & 0.00 & 0.00 & \textbf{0.99} & 0.01 & 0.00 \\   
HUN &   \textBF{0.85} & 0.15 & 0.65 & 0.35 & 0.01 & 0.06 & 0.04 & 0.29 & 0.31 & 0.29 \\
ISL  &  \textbf{0.79}  &  0.21  & 0.60 & 0.40 & \textbf{1.00} & 0.00 & 0.00 & 0.00 & 0.00 & 0.00 \\      
IRL & \textBF{0.97}  &  0.03 & \textBF{0.88} & 0.12 & 0.01 & 0.02 & 0.10 & 0.20 & 0.06 & 0.61 \\    
ITA &  \textBF{0.95}  &  0.05 & \textBF{0.88} & 0.12 & 0.01 & 0.01 & 0.03 & 0.09 & \textbf{0.76} & 0.10 \\
JPN  & 0.24  & \textbf{0.76} & \textBF{0.84} & 0.16 & 0.01 & \textbf{0.91} & 0.02 & 0.03 & 0.02 & 0.01 \\ 
LVA    &  0.01 & \textBF{0.99} & 0.02 & \textBF{0.98} & 0.01 & 0.04 & \textbf{0.90} & 0.01 & 0.03 & 0.01 \\
LTU  & 0.06 & \textBF{0.94} & 0.05 & \textbf{0.95}  & 0.01 & 0.03 & \textbf{0.92} & 0.02 & 0.01 & 0.01 \\
LUX  &  \textBF{0.93} & 0.07 & \textBF{0.96} & 0.04 & 0.01 & 0.03 & 0.07 & 0.21 & 0.63 & 0.05 \\
NLD      & \textBF{0.96} & 0.04 & \textBF{0.99} & 0.01  & 0.01 & 0.10 & 0.11 & 0.16 & 0.61 & 0.01 \\
NZL  & \textBF{0.98} & 0.02 & \textBF{0.96} & 0.04 & 0.01 & 0.03 & 0.36 & 0.37 & 0.14 & 0.09 \\
NOR  & 0.57 & 0.43 & 0.35 & 0.65  & 0.01 & 0.07 & 0.15 & 0.69 & 0.07 & 0.01 \\
POL     & \textBF{0.96} & 0.04  & 0.23 & \textbf{0.77} & 0.01 & 0.01 & 0.01 & \textbf{0.92} & 0.02 & 0.03 \\
PRT    & \textBF{0.99} & 0.01    & \textBF{0.98} & 0.02 & 0.01 & 0.01 & 0.03 & \textbf{0.91} & 0.03 & 0.01 \\
RUS   & 0.03 & \textBF{0.97} & 0.09 & \textbf{0.91} & 0.01 & \textbf{0.93} & 0.02 & 0.01 & 0.02 & 0.01 \\
SVK   & 0.48 & 0.52  & 0.29 & \textBF{0.71} & 0.01 & 0.03 & \textbf{0.82} & 0.04 & 0.09 & 0.01 \\
ESP  & \textBF{0.99} & 0.01 & \textBF{0.82} & 0.18 & 0.01 & 0.07 & 0.08 & \textbf{0.76} & 0.07 & 0.01 \\
SWE    & \textBF{0.98} & 0.02 & \textBF{0.96} & 0.04 & 0.01 & 0.01 & 0.11 & 0.03 & \textbf{0.83} & 0.01 \\
CHE  & \textBF{0.95} & 0.05 & \textBF{0.97} & 0.03 & 0.01 & 0.01 & \textbf{0.85} & 0.06 & 0.05 & 0.02 \\
TWN & \textBF{0.99} & 0.01   & \textBF{0.97} & 0.03 & 0.01 & 0.02 & 0.02 & 0.05 & \textbf{0.89} & 0.01 \\ 
GBR   & \textBF{0.98} & 0.02 & \textBF{0.99} & 0.01 & 0.01 & 0.01 & 0.01 & \textbf{0.73} & 0.14 & 0.10  \\
UKR    & 0.03 & \textBF{0.97} & 0.02 & \textBF{0.98} & 0.01 & \textbf{0.91} & 0.03 & 0.02 & 0.02 & 0.01 \\ 
USA   & 0.02 & \textBF{0.98}  & 0.04 & \textBF{0.96} & 0.01 & \textbf{0.90} & 0.06 & 0.01 & 0.01 & 0.01 \\
\bottomrule
\end{longtable}
\end{center}}

As in the first application, we decided to summarize the characteristics of the countries in each group for the fuzzy $C$-medoids model via weighted averages of the corresponding FQA-based features. These quantities are given in the top part of Table \ref{wavgcountries}. 

\begin{table}[!htb]
\centering
\tabcolsep 0.235in
\caption{Values of $\overline{\rho}_c(\tau_1, \tau_2, 1, \tau_1, \tau_2)$ for two-cluster solutions produced by the fuzzy $C$-medoids procedure based on $\widehat{d}_{\text{FQA}}$ with the 33 functional time series of mortality improvement rates. The top and bottom parts are associated with the total and male populations, respectively.}\label{wavgcountries}
\begin{tabular}{@{}lrrrrrrrr@{}} 
\toprule
Total & \multicolumn{3}{c}{$\tau_2$} & & & \multicolumn{3}{c}{$\tau_2$} \\
$\mathcal{C}_1$ & 0.1 & 0.5 & 0.9 & & $\mathcal{C}_2$ & 0.1 & 0.5 & 0.9 \\\midrule 
$\tau_1=0.1$ & -0.06 & -0.14 & -0.17 & & $\tau_1=0.1$ & 0.05 & 0.01 & -0.04 \\
$\tau_1=0.5$ & -0.12 & -0.18 & -0.15 & & $\tau_1=0.5$ & 0.00 & 0.03 & 0.01 \\
$\tau_1=0.9$ & -0.13 & -0.14 & -0.06 & & $\tau_1=0.9$ & -0.02 & 0.05 & 0.12 \\\midrule
Male & \multicolumn{3}{c}{$\tau_2$} & & & \multicolumn{3}{c}{$\tau_2$} \\
$\mathcal{C}_1$ & 0.1 & 0.5 & 0.9 & & $\mathcal{C}_2$ & 0.1 & 0.5 & 0.9 \\\midrule 
$\tau_1=0.1$ & -0.03 & -0.09 & -0.12 & & $\tau_1=0.1$ & 0.30 & -0.02 & -0.40 \\
$\tau_1=0.5$ & -0.07 & -0.08 & -0.08 & & $\tau_1=0.5$ & -0.05 & -0.19 & -0.10 \\
$\tau_1=0.9$ & -0.11 & -0.05 & -0.03 & & $\tau_1=0.9$ & -0.42 & -0.16 & 0.32 \\\bottomrule
\end{tabular}
\end{table}

Cluster $\mathcal{C}_1$ is associated with higher absolute values in almost all cases, thus indicating that the mortality improvement rates of the countries in this cluster show a stronger degree of serial dependence. On the other hand, most values for cluster $\mathcal{C}_2$ are close to zero. Therefore, the series in this cluster could be assumed to be generated from a stochastic process that is close to an i.i.d. process, thus indicating, for instance, that the yearly curves of mortality improvement rates for the corresponding countries are difficult to predict from past information. 

As an additional exploratory exercise, we decided to repeat the above analyses considering the 33 functional time series of mortality improvement rates for the male and female populations separately. The hyperparameter selection procedure described in Section 3.2 in the main text resulted in $\mathcal{L}=\{1\}$, $C=2$ and $m=1.5$ for the male data and in $\mathcal{L}=\{1, 2\}$, $C=6$ and $m=1.3$ for the female data. 

The clustering partitions associated with the male and female populations for the fuzzy $C$-medoids model are given in the middle and right parts of Table~\ref{tablefuzzycountries}, respectively. Note that even though the clustering solution for the male population involves the same number of groups as that for the total population, both partitions are substantially different. In fact, the solution for the male population includes a cluster $\mathcal{C}_1$ containing all the countries except for ISL, which is the medoid of cluster $\mathcal{C}_2$, thus indicating that the male mortality data of this country are dramatically different from the rest in terms of serial dependence patterns. On the other hand, the solution for the female population is quite difficult to interpret since it involves six groups. {The fuzzy solutions for the male and female populations regarding the fuzzy $C$-means model are given in the middle and right parts of Table~\ref{tablefuzzycountries2}, respectively.}

The weighted averages of the corresponding FQA-based features for the male and female data associated with the fuzzy $C$-medoids model are provided in the bottom part of Table~\ref{wavgcountries} and Table~\ref{wavgcountriesf}, respectively. Note that, for the male population, some quantities associated with cluster $\mathcal{C}_2$ are quite large in absolute value, implying that the male mortality data of ISL exhibit a substantially stronger degree of serial dependence than the ones displayed by its counterparts. On the other hand, for the female population, the six clusters are associated with different dependence patterns, and, in all cases, the groups display a stronger degree of dependence at the first lag ($l=1$) than at the second lag ($l=2$).   


\begin{table}[!htbp]
\centering
\renewcommand*{\arraystretch}{1}
\tabcolsep 0.23in
\caption{Values of $\overline{\rho}_c(\tau_1, \tau_2, 1, \tau_1, \tau_2)$ and $\overline{\rho}_c(\tau_1, \tau_2, 2, \tau_1, \tau_2)$ for the six-cluster solution produced by the fuzzy $C$-medoids procedure based on $\widehat{d}_{\text{FQA}}$ with the 33 functional time series of mortality improvement rates for the female population.}\label{wavgcountriesf}
\begin{tabular}{@{}lrrrrrrrr@{}} 
\toprule
  $l=1$  & \multicolumn{3}{c}{$\tau_2$} & & $l=2$ & \multicolumn{3}{c}{$\tau_2$} \\
$\mathcal{C}_1$   & 0.1 & 0.5 & 0.9 & &  $\mathcal{C}_1$ & 0.1 & 0.5 & 0.9  \\ \midrule 
$\tau_1=0.1$  &  -0.06                    &        -0.20 & -0.21 & &  $\tau_1=0.1$ & -0.01   & 0.00 & -0.03                  \\
$\tau_1=0.5$    & -0.18                      &   -0.25 &  -0.16 & & $\tau_1=0.5$     &      -0.04        & -0.04 &   -0.05       \\
$\tau_1=0.9$  & -0.20                        &   -0.20   & -0.09 & & $\tau_1=0.9$      & -0.04 & -0.02 &  -0.05             \\ \hline 
    & \multicolumn{3}{c}{$\tau_2$} & & & \multicolumn{3}{c}{$\tau_2$} \\
$\mathcal{C}_2$   & 0.1 & 0.5 & 0.9 & &  $\mathcal{C}_2$ & 0.1 & 0.5 & 0.9  \\ \midrule 
$\tau_1=0.1$  &  -0.04                    &        -0.10 & -0.03 & &  $\tau_1=0.1$ & 0.02   & 0.02 & -0.02                  \\
$\tau_1=0.5$    & -0.08                      &   -0.09 &  0.00 & & $\tau_1=0.5$     &      0.06        & 0.07 &   0.00       \\
$\tau_1=0.9$  & -0.17                        &   -0.07   & 0.00 & & $\tau_1=0.9$      & 0.03 & 0.06 &  -0.04             \\ \hline 
    & \multicolumn{3}{c}{$\tau_2$} & & & \multicolumn{3}{c}{$\tau_2$} \\
$\mathcal{C}_3$   & 0.1 & 0.5 & 0.9 & &  $\mathcal{C}_3$ & 0.1 & 0.5 & 0.9  \\ \midrule 
$\tau_1=0.1$  &  -0.02                    &        -0.13 & -0.11 & &  $\tau_1=0.1$ & 0.02   & 0.00 & 0.01                  \\
$\tau_1=0.5$    & -0.17                      &   -0.25 &  -0.12 & & $\tau_1=0.5$     &      -0.03        & 0.01 &   -0.03       \\
$\tau_1=0.9$  & -0.30                        &   -0.26   & -0.09 & & $\tau_1=0.9$      & -0.02 & 0.05 &  -0.03             \\ \hline 
        & \multicolumn{3}{c}{$\tau_2$} & & & \multicolumn{3}{c}{$\tau_2$} \\
$\mathcal{C}_4$   & 0.1 & 0.5 & 0.9 & &  $\mathcal{C}_4$ & 0.1 & 0.5 & 0.9  \\ \midrule 
$\tau_1=0.1$  &  0.00                    &        -0.18 & -0.14 & &  $\tau_1=0.1$ & 0.15   & 0.03 & -0.08                  \\
$\tau_1=0.5$    & -0.12                      &   -0.25 &  -0.10 & & $\tau_1=0.5$     &      0.10        & 0.04 &   -0.14       \\
$\tau_1=0.9$  & -0.22                        &   -0.19   & -0.05 & & $\tau_1=0.9$      & 0.13 & 0.12 &  -0.05             \\ \hline 
        & \multicolumn{3}{c}{$\tau_2$} & & & \multicolumn{3}{c}{$\tau_2$} \\
$\mathcal{C}_5$   & 0.1 & 0.5 & 0.9 & &  $\mathcal{C}_5$ & 0.1 & 0.5 & 0.9  \\ \midrule 
$\tau_1=0.1$  &  -0.08                    &        -0.18 & -0.23 & &  $\tau_1=0.1$ & 0.15   & 0.11 & 0.02                  \\
$\tau_1=0.5$    & -0.16                      &   -0.24 &  -0.18 & & $\tau_1=0.5$     &      0.06        & 0.02 &   -0.04       \\
$\tau_1=0.9$  & -0.20                        &   -0.18   & -0.03 & & $\tau_1=0.9$      & -0.02 & 0.00 &  0.02             \\ \hline 
        & \multicolumn{3}{c}{$\tau_2$} & & & \multicolumn{3}{c}{$\tau_2$} \\
$\mathcal{C}_6$   & 0.1 & 0.5 & 0.9 & &  $\mathcal{C}_6$ & 0.1 & 0.5 & 0.9  \\ \midrule 
$\tau_1=0.1$  &  -0.11                    &        -0.18 & -0.23 & &  $\tau_1=0.1$ & 0.18   & 0.09 & 0.04                  \\
$\tau_1=0.5$    & -0.27                      &   -0.28 &  -0.22 & & $\tau_1=0.5$     &      0.08        & 0.06 &   0.11       \\
$\tau_1=0.9$  & -0.30                        &   -0.21   & -0.09 & & $\tau_1=0.9$      & 0.00 & 0.06 &  0.17             \\ \hline 
\end{tabular}
\end{table}

\newpage
\bibliographystyle{agsm}
\bibliography{FTS_clustering}
	
\end{document}